**Can Large Language Models Represent Urban Publics? Behavioral Replication and Population Mismatch in an Affordable-Housing Experiment**

Yuxuan Cai [a, *], Yequan Hu [b, c], Hongqian Li [d, e], Zhanghong Ju [f], and Shuying Guo [g]

a. *Department of City and Regional Planning, The University of North Carolina at Chapel Hill, Chapel Hill, NC, USA,* yuxuanc@unc.edu

b. *School of Geographical Sciences, University of Bristol, Bristol BS8 1SS, UK,* zczqhuj@ucl.ac.uk

c. *AML Lab, College of Computing, City University of Hong Kong, Hong Kong SAR, 999077, China*

d. *Graduate School of Architecture, Planning and Preservation, Columbia University, NY, USA, hl3741@columbia.edu*

e. *Department of Geography, UC Santa Barbara, Santa Barbara, CA, USA, hongqian@ucsb.edu*

f. *Department of City and Regional Planning, The University of North Carolina at Chapel Hill, Chapel Hill, NC, USA, juz@unc.edu*

g. *Department of Human Geography and Spatial Planning, Faculty of Geosciences, Utrecht University, Utrecht, 3584, CB, the Netherlands, s.guo1@uu.nl*

* *Correspondence: yuxuanc@unc.edu*

**Abstract**

There is growing interest in using large language models (LLMs) as low-cost proxies for resident attitudes in urban planning. Previous work shows that LLMs can predict average results of survey experiments, but less is known about whether they preserve the spatially anchored, identity-conditioned structure behind those averages, namely how support changes as a project approaches homes and how that response divides across tenure and partisan groups. We compared eight open-weight LLMs with 843 respondents in a US affordable-housing survey experiment, testing whether they reproduced the owner–renter difference in support change as a proposed development moved from 2 miles to 1/8 mile. Qwen 2.5 14B was closest (−0.242 versus the human −0.285) and was the only model to meet the prespecified ±0.20 equivalence criterion; Phi-4 14B was directionally aligned but attenuated (−0.150), and other models showed weak, null, or reversed moderation. This aggregate match masked structural failure. Qwen attenuated the Republican contrast and exaggerated the Independent one, its RMSE across 27 party-by-tenure-by-item cells was 0.613, its median model-to-human variance ratio was 0.099, and question order shifted the contrast by +0.367. Identity-cue removal and selective nonresponse changed which comparisons were estimable, and rationale-first responses differed from matched direct-choice responses in 20.6–35.3% of focal comparisons. An LLM can thus approximate one aggregate contrast while failing to preserve the population structure, within-group heterogeneity, and measurement stability that generate it. Model evaluation in urban planning should test whether this spatial and social structure survives simulation, not only average effects.

**Keywords:** urban attitudes; spatial proximity; synthetic respondents; large language models

## 1 Introduction

As participatory planning has matured, residents' attitudes toward development at specific locations have become a formal input to the planning process (Innes and Booher, 2004). Planning decisions about housing, infrastructure, and neighborhood change depend not only on average public support but also on whose interests are represented and how attitudes vary with spatial exposure. Affordable-housing siting makes this dependence visible. Moving a proposed development closer to home can change perceived local costs and benefits. At the same time, housing tenure and political identity organize residents' relationships to property values, rents, neighborhood change, and public provision (Anderson et al., 2026; Hankinson, 2018; Marble and Nall, 2021).

In affordable housing studies, a substantial literature has taken these locational attitudes as its object of study, and cross-sectional surveys have mapped who opposes new and affordable housing and why, linking opposition above all to political ideology and to stereotypes about prospective residents' race and class (Tighe, 2012; Whittemore and BenDor, 2019). Survey experiments have identified the causal effects of proximity and project attributes (Hankinson, 2018), while related experiments on market-rate development show that how a proposal is framed independently moves support (Monkkonen and Manville, 2019). Although these methods reach samples of the broader public, the projects are hypothetical. Analyses of public-meeting records, on the other hand, have captured participation as actual behavior, whereas the records only rely on the residents who show up (Einstein et al., 2019). Under these constraints, predicting how a specific group responds to a given site at scale remains unattainable, as every new site or subgroup requires fielding a new sample. A computational measure of urban opinion is therefore useful only if it preserves the spatial and social structure of the population it is intended to represent.

Given a demographic profile, a large language model (LLM) can generate responses for any group and any site without fielding a new sample (Aher et al., 2023; Argyle et al., 2023; Hu and Collier, 2024). Recent high-profile studies show both sides of this prospect: models can predict many aggregate treatment effects, yet they may systematically overestimate effect sizes (Ashokkumar et al., 2026), flatten identity-group variation (Bisbee et al., 2024; Santurkar et al., 2023; Wang et al., 2025), or change with framing (Bisbee et al., 2024; Brucks and Toubia, 2025; Tjuatja et al., 2024). For urban research, these are not abstract model-quality defects. A planning estimate can appear accurate in aggregate while misrepresenting the constituencies that bear different spatial costs and benefits.

The unresolved gap lies in the unit of validation. Existing evaluations often ask whether a model recovers an overall mean, correlation, or recognizable behavioral range (Ashokkumar et al., 2026; Mei et al., 2024). Urban policy questions instead require population-conditional comparisons: a pooled owner-renter contrast may look correct while one political group is exaggerated, another is attenuated, and within-group disagreement is collapsed. Question order, identity cues, preceding context, and requests for justification may further change the measured response (Lutz et al., 2025; Turpin et al., 2023). The relevant planning question is therefore not only whether a model matches an effect but also whether it preserves the comparison structure needed to describe affected populations. Recent urban studies have compared AI and human judgments across residents and non-residents or national and local samples (Malekzadeh et al., 2025; Bollen et al., 2026). Our design addresses a distinct validation problem:

whether a model reproduces a randomized spatial-proximity effect and the party-by-tenure distributions, order stability, and estimability structure that generate it.

We address this question using a completed U.S. affordable-housing survey experiment as an external human benchmark. Eight open-weight models are first evaluated on the ownership-conditioned proximity effect and the subgroup cells, response distributions, and randomized-order contrasts that produce it. We then test how identity-cue removal, prompt framing, preceding context, and model-specific nonresponse affect what remains estimable. Finally, in three 12–14B systems, we compare matched response formats and apply a single fixed exploratory hidden-state diagnostic.

The study contributes a population-aware framework for evaluating AI-generated urban attitudes. It separates four questions that are often collapsed: whether a model reproduces a target spatial effect, whether the subgroup distributions generating that effect are preserved, whether the result survives changes in information and elicitation, and whether explanation-oriented diagnostics support any shared representation. This design allows apparent success and consequential failure to be evaluated within the same experiment rather than treated as competing conclusions. For computational urban systems, it treats an AI-generated attitude as a spatially indexed measurement output whose validity depends jointly on the exposure contrast and the population represented.

## 2 From spatial exposure to population-conditional urban attitudes

Affordable-housing attitudes are relational rather than context-free. Moving the same proposal closer to home changes the local stakes, but those stakes are not uniform: tenure changes how residents relate to property values and rents, while political identity changes how public provision and neighborhood change are interpreted (Anderson et al., 2026; Hankinson, 2018; Marble and Nall, 2021). The human benchmark is therefore not a simple near-far mean. It is a contrast of spatial responses across tenure groups, evaluated within a politically heterogeneous population. We assess LLM-generated attitudes at three linked levels: spatial-behavioral correspondence, population-structural correspondence, and measurement stability. The response-format experiment and hidden-state RSA provide separate exploration diagnostics.

RQ1 (spatial-behavioral correspondence): Do open-weight LLMs reproduce the human owner–renter difference in the support penalty caused by moving an affordable-housing project closer to home? RQ2 (population-structural correspondence): When a model is close on that contrast, does it also preserve party-specific moderation, complete response distributions, dispersion, cell-level means, and randomized-order stability? RQ3 (measurement and information dependence): Which conclusions remain identifiable when party and tenure cues are removed, prompts are reframed, preceding context is added, or valid responses are unevenly distributed across identity groups? RQ4 (exploratory explanation diagnostic): Do matched response-format conditions alter choices, and does a fixed hidden-state RSA yield a shared attitude-specific representation pattern across models?

## 3 Methods

### 3.1 Study architecture

We compared responses from eight open-weight LLMs with those of 843 participants in a U.S. survey experiment on affordable housing. Our primary estimand was the owner–renter difference in the distance effect, defined as the change in support when the proposed development was described as 1/8 mile rather than 2 miles away. For each model, we compared this estimate with the corresponding estimate in the human sample. Secondary analyses estimated the same contrast within party-by-tenure subgroups, compared the full model and human response distributions, and tested whether the estimate varied with question order (Fig. 1 and Fig. 2).

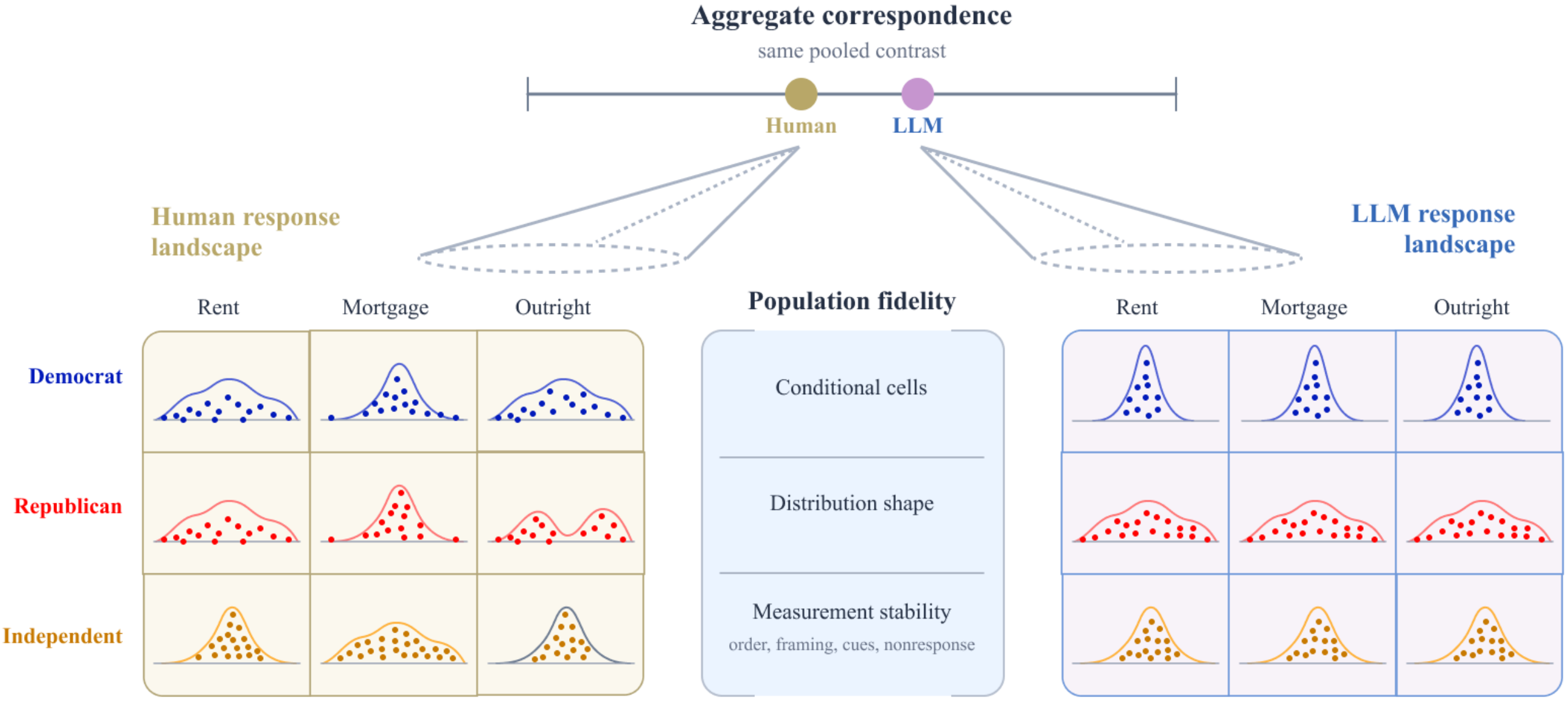

Fig. 1. Conceptual framework for evaluating behavioral correspondence, population structure and measurement stability in LLM-generated urban attitudes. The distributions shown are schematic.

We next assessed prompt sensitivity. We ran Qwen and Mistral in three experiments that varied identity cues (four conditions), task framing (three), and preceding context (four), respectively. For Phi-4 and Gemma 3, the identity-cue analysis was limited to a one-attempt 32-session coverage screen spanning 16 visible profiles and two question orders. Outputs that failed the prespecified response schema were classified as incomplete and retained for the nonresponse audit.

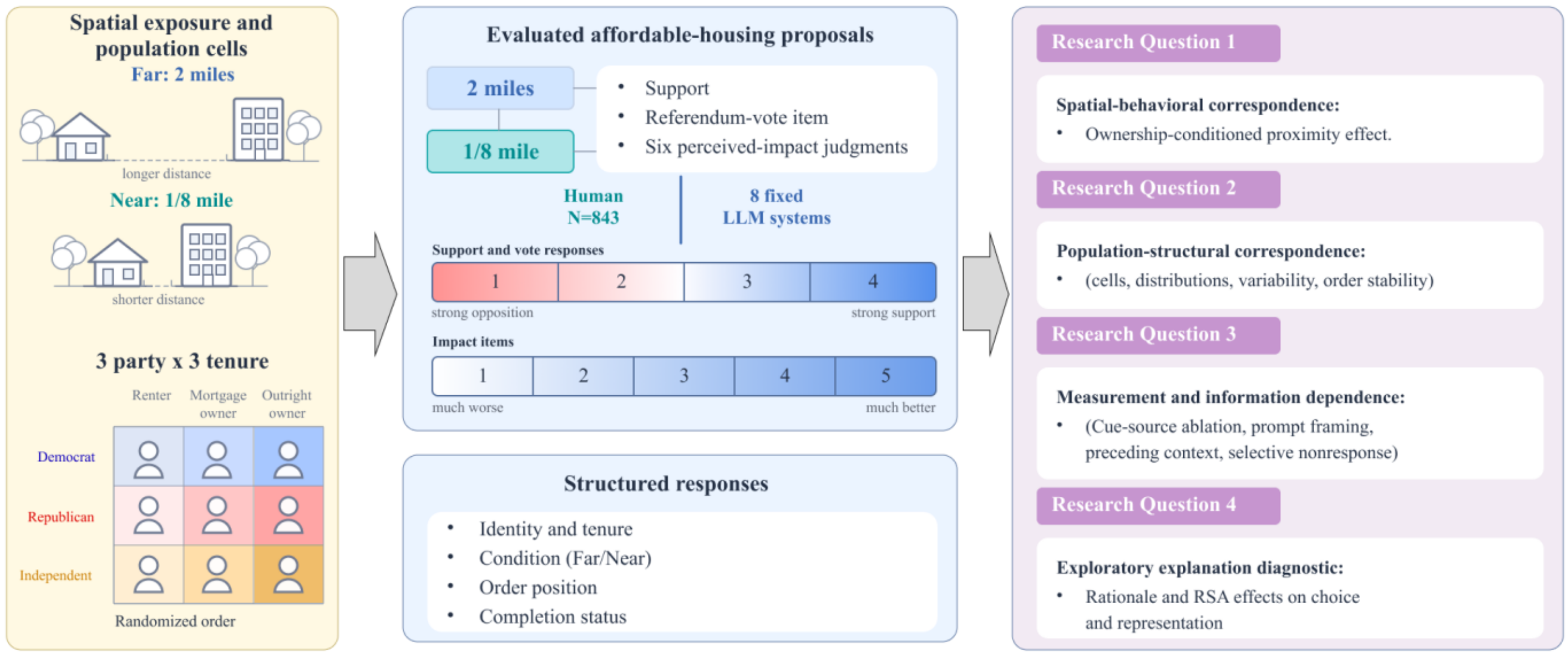

Fig. 2. Study framework.

In a separate response-format experiment, Qwen 2.5 14B, Phi-4 14B, and Gemma 3 12B answered matched prompts in three formats, which included choice only, choice followed by a brief explanation, and a brief explanation followed by choice. As an exploratory analysis, we applied representational similarity analysis to hidden states from the same three models. The sections below define the analytic sample and estimand for each analysis.

### 3.2 Human benchmark and experimental design

The human benchmark was the control condition of Anderson et al.'s U.S. affordable-housing survey experiment (Anderson et al., 2026). The original study recruited 3,001 U.S. adults through CloudResearch Connect in May 2024. Respondents were randomized across intervention conditions and local-question order, with demographic quotas and later Republican oversampling. The primary Human benchmark comprised 843 control-condition records with complete primary outcomes, each assignable to one of the nine prespecified party-by-tenure cells. As a sensitivity analysis, we re-estimated the Human benchmark after removing 10 records associated with the source study's initial soft launch (N = 833; Supplementary Section S7). In the primary 843-record benchmark, the cells contained 112, 102 and 66 Democratic renters, mortgage holders, and outright owners; 115, 157, and 88 Republican respondents; and 83, 63 and 57 Independent respondents, respectively.

Each respondent evaluated affordable-housing proposals located 1/8 mile and 2 miles from home. The two local modules were presented in randomized near-first or far-first order. Each module included support, a referendum-vote item, and six perceived-impact judgments. Support and vote responses were scored from 1 to 4, with higher values indicating greater support. Impact items were scored from 1 to 5, with higher values indicating a more positive expected impact. The present study uses human data as a benchmark, not as evidence that model sessions are drawn from the same population.

### 3.3 Model systems, generation, and valid completion

Study 1 evaluates Qwen/Qwen2.5-14B-Instruct, microsoft/phi-4, google/gemma-3-12b-it, Google/Gemma-2-9b-it, Meta-llama/Llama-3.1-8B-Instruct, Qwen/Qwen2.5-7B-Instruct, Allenai/OLMo-2-1124-7B-Instruct, and mistralai/Mistral-7B-Instruct-v0.3 (Abdin et al., 2024; Gemma Team et al., 2025, 2024; Grattafiori et al., 2024; Jiang et al., 2023; OLMo et al., 2025; Qwen et al., 2024). The original six-system benchmark used NF4 four-bit weights with double quantization and float16 computation (Dettmers et al., 2023). The Phi-4 and Gemma 3 extensions used unquantized bfloat16 computation on A100 GPUs. Temperature (0.7), top-p (0.9), maximum input length (8,192), response schema, questionnaire content, scoring, and base sampling design were held fixed. Numerical precision is therefore documented as part of the system definition; cross-system differences are descriptive and are not attributed causally to architecture, family, or parameter count.

For each model, the intended main design crossed three party identities (Democrat, Republican, Independent), three tenure states (rent, mortgage, outright ownership), and two randomized local-module orders. Each of the 18 strata contained 40 sessions, yielding 720 intended sessions per model. A session is an independent pseudo-random generation under a fixed model and decoder, not a simulated real person.

### 3.4 Study 1: behavioral and structural fidelity

#### 3.4.1 Primary behavioral estimand

We first measured how much support changed when the proposal moved closer to a respondent's home. For each human respondent or model session s, we subtracted the 2-mile support score from the 1/8-mile score:

$$D_s = \text{support}_{1/8,s} - \text{support}_{2mi,s} \quad (1)$$

Negative values of $D_s$ indicate lower support for the nearby proposal.

We next asked whether proximity changed support more for owners than renters. Within each party p, we averaged the mortgage and outright-owner proximity effects and subtracted the renter effect.

$$\Delta_p = 0.5(\bar{D}_{p,\text{mortgage}} + \bar{D}_{p,\text{outright}}) - \bar{D}_{p,\text{rent}} \quad (2)$$

where $\bar{D}_{p, t}$ denotes the mean of the respondent- or session-level proximity effects within party p and tenure group t. The primary estimand averaged these contrasts across the three party groups:

$$\theta = (\Delta\text{Democrat} + \Delta\text{Republican} + \Delta\text{Independent}) / 3$$

Each party therefore contributed equally to θ, irrespective of its sample size. A more negative value indicates that owners, relative to renters, showed a larger decline in support when the proposal was nearby.

We obtained 95% confidence intervals for θ from 2,000 bootstrap resamples. Human respondents and model sessions were resampled separately within each party-by-tenure cell. To assess similarity with the human benchmark on this estimand, we calculated the difference between each model estimate and the human estimate in every bootstrap resample. Practical equivalence was defined as the resulting 90% interval falling entirely within a study-specific margin of ±0.20 support points (Lakens et al., 2018).

Sensitivity to alternative margins of ±0.10, ±0.15 and ±0.25 is reported in Supplementary Section S1 and Supplementary Table S1.

### 3.4.2 Structural and distributional fidelity

Similarity in the pooled value of θ does not necessarily imply similarity across the population cells from which it is calculated. We therefore examined the party-specific owner–renter contrasts and the proximity effect within each of the nine party-by-tenure cells.

We also compared the human and model support-response distributions across 27 cells formed by crossing three support items (statewide, 1/8 mile, and 2 miles), three party groups, and three tenure groups.

We used total variation and Jensen–Shannon distances to characterize overall distributional separation. Ordinal Wasserstein distance additionally accounted for the ordering of the four response categories. Variance ratios and differences in normalized entropy relative to the human benchmark were used to assess whether model responses were compressed within cells. Pearson and Spearman correlations assessed correspondence in cell means, whereas mean absolute error and root mean squared error measured their absolute deviation.

We conducted two additional sensitivity analyses to address the unequal composition of the human sample. First, we recalculated the primary contrast for the human benchmark and each model using weights derived from the observed human party shares and the observed composition of mortgage and outright owners within each party. Second, we compared human–model distribution distances with a matched-sample human–human bootstrap baseline. This baseline quantified the distance expected between two finite samples drawn from the same empirical human response distributions. Full definitions and uncertainty calculations are reported in Supplementary Section S2, Supplementary Tables S2–S4, and Supplementary Fig. S1.

Finally, we examined question-order sensitivity by estimating the overall proximity effect and θ separately for near-first and far-first sessions. The order shift was calculated as the near-first estimate minus the far-first estimate. Bootstrap resampling retained all 18 party-by-tenure-by-order strata, and Holm correction was applied within each family of order comparisons.

## 3.5 Study 2: information, framing, context, and nonresponse

We assessed the dependence of model responses on prompt information through three experiments. First, we removed party and tenure cues separately and jointly while leaving the housing scenarios unchanged. This experiment compared responses generated with both identity cues, either cue alone, and neither cue. Second, we presented the same respondent profiles and housing questions under three task descriptions: responding as the described respondent, predicting that respondent's likely answer, and completing a survey record. Third, we changed the questions preceding the focal housing items by presenting no additional questions, the original survey's ten-item pretreatment block (Anderson et al., 2026), a no-feedback block consisting of an instruction acknowledgement and eight true-or-false affordable-housing questions, or both blocks. We applied all three designs to Qwen 2.5 14B and Mistral 7B. Phi-4 14B and Gemma 3 12B were examined only in the smaller identity-cue coverage screen. Together, these experiments distinguish dependence on respondent information from dependence on task description and preceding questionnaire content.

In the identity-cue experiment, prompts included both party and tenure, party only, tenure only or neither attribute. The housing scenarios and response questions remained unchanged across conditions. We assessed whether the response observed when both cues were present differed from what would be expected from their separate associations, using the no-cue condition as the baseline. Because the single-cue conditions did not define complete party-by-tenure profiles, these comparisons averaged equally across the identity profiles observable in each condition rather than applying the primary estimand θ.

The three task-framing conditions held respondent information and housing questions constant and differed only in the instruction used to elicit a response. The model was instructed to continue as the respondent, predict one plausible response, or complete a survey record.

In the preceding-context experiment, the focal housing items followed no additional questions, the original survey's ten-item pretreatment block, the no-feedback affordable-housing knowledge block, or both blocks. The respondent profile and the order of the two local housing modules were held constant across conditions.

Responses that could not be assigned unambiguously to a listed option were classified as nonresponses. Completion, estimability, and the bounds obtained when required profiles were unobserved are reported in Supplementary Section S3 and Supplementary Table S5.

### 3.6 Study 3: response-format sensitivity and exploratory hidden-state RSA

We conducted two analyses in Study 3. The first tested whether asking models to explain an answer before or after selecting it changed their observable responses. The second used exploratory representational similarity analysis (Kriegeskorte, 2008) to test whether pairwise dissimilarities among model hidden states tracked physical distance in housing-attitude prompts beyond the structure observed for literal distance-token control. The two analyses address response-format sensitivity and internal representation, respectively; neither was interpreted as evidence of human-like reasoning.

We evaluated Qwen 2.5 14B, Phi-4 14B, and Gemma 3 12B under three matched response formats: choice only, choice followed by a brief explanation, and a brief explanation followed by choice. For each model, five prompt sets were selected within each of the 18 party-by-tenure-by-order strata, producing 90 matched prompt sets. Each set was generated under all three response formats, yielding 270 model runs per model. Respondent profiles, housing questions, and question order were held constant across formats, while each run remained an independent generation. We compared each explanation condition with its matched choice-only condition across the four near and far support and vote items, yielding 360 item-level comparisons per explanation condition. We report exact agreement, mean item changes, and differences in the near-minus-far outcomes, with 95% confidence intervals obtained from 2,000 bootstrap resamples of prompt sets within the 18 strata.

We used a prespecified keyword dictionary to describe the content of the generated explanations. The dictionary identified references to 11 categories: distance or proximity, housing tenure, self-interest, property values, taxes or public costs, housing supply or affordability, social welfare, neighborhood quality, political identity, fairness, and uncertainty. Multiple categories could be assigned to the same explanation. These dictionary-based indicators were interpreted as observable features of the generated text, not as measures of latent reasoning or psychological mediation.

To examine whether internal model representations reflected physical distance, we analyzed a common set of 128 prompts for each model. The design combined four distances (0.125, 2, 10 and 50 miles), four prompt categories, and eight paraphrases. The categories covered housing attitudes, housing knowledge, non-housing spatial questions, and literal distance-token controls. We focused on hidden states at 75% of model depth and averaged the representations of tokens specifying distance. For each prompt category, we summarized the relationships among the four distances using cosine dissimilarity. We then tested whether this representational pattern followed the rank ordering of pairwise log-distance differences. The primary diagnostic compared this correlation between housing-attitude prompts and literal distance-token controls. This contrast was designed to distinguish attitude-related spatial structure from the representation of distance terms alone. We quantified uncertainty using 2,000 bootstrap resamples and assessed statistical significance using 5,000 permutations. Analyses at other layers and with alternative pooling rules served as sensitivity tests, with Holm correction applied within each model and diagnostic family.

## 4 Results

### 4.1 Primary behavioral test

To evaluate whether any model reproduced the human proximity pattern, we first estimated the overall proximity effect and the prespecified ownership-conditioned contrast $\theta$, in the human benchmark.

Moving the proposed project from 2 miles to 1/8 mile reduced mean support by 0.465 points, and this decline was larger among owners than renters ($\theta = -0.285$, 95% CI [−0.385, −0.179]; Table 1 and Fig. 3). Only Qwen 2.5 14B met the prespecified equivalence criterion for $\theta$, with an estimate of −0.242 (95% CI [−0.302, −0.183]) and a model–human difference of +0.044 (90% CI [−0.058, +0.139]), entirely within the ±0.20 equivalence region. Despite meeting the equivalence criterion for $\theta$, Qwen produced a larger overall proximity effect than humans (−0.644 versus −0.465). Across the remaining systems, the direction and magnitude of the two proximity estimates varied.

Phi-4 showed a modest decline in support as the project moved closer (−0.271), together with a weaker owner–renter contrast in the same direction as the human benchmark (−0.150). By contrast, Gemma 3 12B and Gemma 2 9B produced substantially larger proximity-related declines (−0.867 and −0.964, respectively) but showed little differentiation between owners and renters (−0.050 and −0.027).

The remaining systems likewise failed to recover the complete human pattern. Llama 3.1 8B reversed the owner–renter contrast (+0.133), whereas the corresponding estimates for Qwen 2.5 7B and OLMo 2 7B were close to zero. Mistral 7B showed increased rather than decreased support for the nearby project (+0.171), although its owner–renter contrast weakly followed the direction observed in humans (−0.069). Thus, none of these systems preserved both the overall response to proximity and its variation by housing tenure.

Performance did not improve consistently with nominal parameter count across the eight systems. However, the systems also differed in model family, training, instruction tuning, and numerical precision.

These descriptive comparisons therefore cannot attribute the observed differences to model scale or capability alone.

To determine whether equal weighting of the three party groups influenced the primary comparison, we re-estimated the contrast using the observed party and owner-type composition of the human benchmark. Reweighting shifted the human estimate from −0.285 to −0.298 and the Qwen estimate from −0.242 to −0.185, increasing their absolute difference from 0.044 to 0.114. Nevertheless, Qwen remained the closest model, and the relative ordering of the eight systems was unchanged. Because this analysis was conducted post hoc, it was interpreted as a sensitivity check rather than as an additional equivalence test (Supplementary Section S2, Supplementary Tables S2–S3 and Supplementary Fig. S1a).

Table 1. Human and model proximity effects.

| Source | N | Overall near−far support | Owner−renter moderation | Interpretation |
|---|---|---|---|---|
| Human | 843 | −0.465 [−0.519, −0.412] | −0.285 [−0.385, −0.179] | Reference pattern |
| Qwen 2.5 14B | 720 | −0.644 [−0.676, −0.614] | −0.242 [−0.302, −0.183] | Closest moderation; stronger overall penalty |
| Phi-4 14B | 720 | −0.271 [−0.296, −0.246] | −0.150 [−0.202, −0.098] | Aligned but attenuated |
| Gemma 3 12B | 719 | −0.867 [−0.888, −0.847] | −0.050 [−0.096, −0.006] | Strong overall; weak tenure differentiation |
| Gemma 2 9B | 720 | −0.964 [−0.976, −0.950] | −0.027 [−0.058, +0.002] | Strong overall; near-absent moderation |
| Llama 3.1 8B | 720 | −1.211 [−1.269, −1.153] | +0.133 [+0.013, +0.250] | Reversed tenure moderation |
| Qwen 2.5 7B | 720 | +0.029 [−0.022, +0.082] | +0.044 [−0.063, +0.154] | Near-zero overall and moderation |
| OLMo 2 7B | 720 | +0.031 [+0.018, +0.043] | +0.008 [−0.017, +0.031] | Near-zero overall and moderation |
| Mistral 7B | 720 | +0.171 [+0.146, +0.197] | −0.069 [−0.121, −0.013] | Overall reversal; weak aligned moderation |

Note. Negative overall values indicate lower support for the near project. More negative moderation indicates a larger near-project penalty among owners than renters.

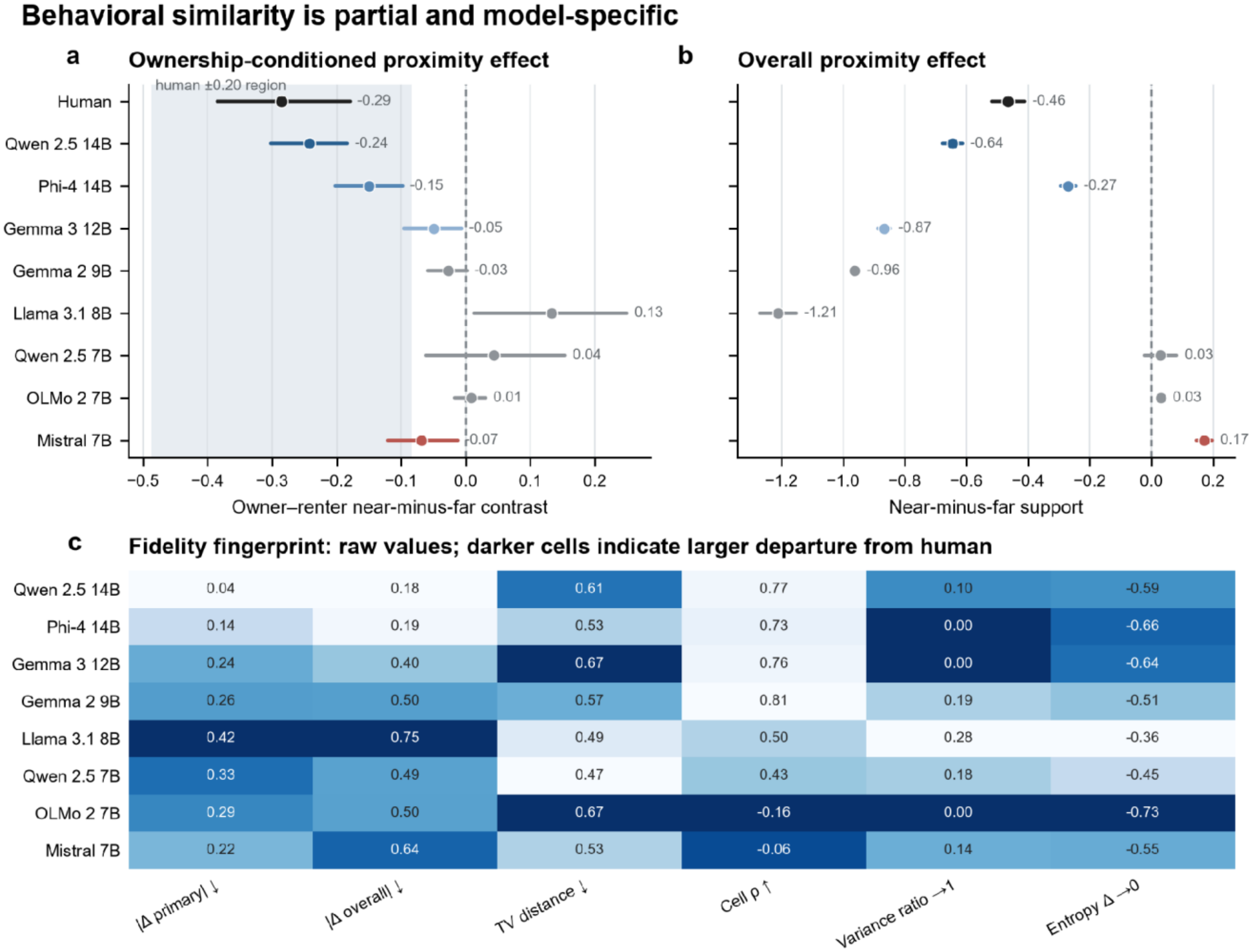

| | \|Δ primary\| ↓ | \|Δ overall\| ↓ | TV distance ↓ | Cell ρ ↑ | Variance ratio →1 | Entropy Δ →0 |
|---|---|---|---|---|---|---|
| Qwen 2.5 14B | 0.04 | 0.18 | 0.61 | 0.77 | 0.10 | -0.59 |
| Phi-4 14B | 0.14 | 0.19 | 0.53 | 0.73 | 0.00 | -0.66 |
| Gemma 3 12B | 0.24 | 0.40 | 0.67 | 0.76 | 0.00 | -0.64 |
| Gemma 2 9B | 0.26 | 0.50 | 0.57 | 0.81 | 0.19 | -0.51 |
| Llama 3.1 8B | 0.42 | 0.75 | 0.49 | 0.50 | 0.28 | -0.36 |
| Qwen 2.5 7B | 0.33 | 0.49 | 0.47 | 0.43 | 0.18 | -0.45 |
| OLMo 2 7B | 0.29 | 0.50 | 0.67 | -0.16 | 0.00 | -0.73 |
| Mistral 7B | 0.22 | 0.64 | 0.53 | -0.06 | 0.14 | -0.55 |

Fig. 3. Behavioral similarity is partial and model-specific. Panel a reports the ownership-conditioned proximity contrast; panel b reports the overall proximity effect; panel c summarizes six complementary fidelity dimensions.

Note. Points and horizontal intervals are estimates and 95% bootstrap intervals. The shaded region in panel a marks the human estimate ±0.20. Darker heatmap cells indicate larger departure after metric-specific orientation.

### 4.2 Subgroup and distributional tests

#### 4.2.1 Party-specific contrasts

For Qwen, the Democratic contrast closely matched the human benchmark (−0.181 versus −0.194). However, Qwen showed much weaker owner–renter differentiation among Republicans (−0.044 versus −0.335) and stronger differentiation among Independents (−0.500 versus −0.326). Because these two errors occurred in opposite directions, they partially cancelled when averaged. This offset explains why Qwen appeared closer to humans on the pooled θ than at the subgroup level.

Phi showed a related pattern. Its Democratic contrast was close to the human benchmark (−0.206 versus −0.194), whereas it substantially underestimated the owner–renter contrasts among Republicans (−0.069 versus −0.335) and Independents (−0.175 versus −0.326). Gemma 3 departed more clearly from the human pattern, showing no ownership moderation among Republicans or Independents. Thus, Qwen, Phi and Gemma 3 each failed to reproduce the human pattern consistently across all three party groups.

### 4.2.2 Response distributions

We next asked whether agreement on θ extended to 27 party-by-tenure-by-item cells (Table 2). Gemma 2 had the highest Pearson correlation with human cell means ($r = 0.796$), although its RMSE was 0.638. Phi had the lowest RMSE (0.384) and a Pearson correlation of 0.747. Qwen's correlation was nearly identical ($r = 0.746$), but its RMSE was higher (0.613). OLMo had an MAE of 0.403, but its cell means were essentially uncorrelated with the human values ($r = -0.011$). Thus, a smaller RMSE or MAE did not necessarily indicate that a model reproduced the human cross-cell pattern.

Table 2. Structural and distributional fidelity across 27 party-by-tenure-by-item cells

| Model | Mean TV | Mean JS | Mean $W_1$ | RMSE | Cell r | Median variance ratio | Entropy Δ |
|---|---|---|---|---|---|---|---|
| Qwen 2.5 14B | 0.615 | 0.279 | 0.771 | 0.613 | 0.746 | 0.099 | −0.590 |
| Phi-4 14B | 0.531 | 0.240 | 0.662 | 0.384 | 0.747 | 0.000 | −0.656 |
| Gemma 3 12B | 0.666 | 0.336 | 0.847 | 0.677 | 0.778 | 0.000 | −0.641 |
| Gemma 2 9B | 0.567 | 0.259 | 0.729 | 0.638 | 0.796 | 0.188 | −0.510 |
| Llama 3.1 8B | 0.495 | 0.192 | 0.691 | 0.661 | 0.607 | 0.281 | −0.362 |
| Qwen 2.5 7B | 0.469 | 0.185 | 0.632 | 0.530 | 0.423 | 0.182 | −0.455 |
| OLMo 2 7B | 0.666 | 0.317 | 0.789 | 0.502 | −0.011 | 0.000 | −0.728 |
| Mistral 7B | 0.532 | 0.231 | 0.758 | 0.706 | −0.113 | 0.141 | −0.549 |

Note. Lower TV, JS, $W_1$, and RMSE are better; correlation is ideally high; variance ratio is ideally 1; entropy difference is ideally 0. The table deliberately retains multiple construct-distinct metrics.

We then compared response variation within the same 27 cells. Qwen's mean total-variation and ordinal Wasserstein distances were 0.615 and 0.771, respectively, and its median model-to-human variance ratio was 0.099. Phi's median variance ratio was 0 despite its low RMSE; Gemma 3 also had a ratio of 0. A median ratio of 0 indicates no model-response variance in at least half of the cells. Entropy was lower than the human benchmark for every model. Cell-mean correspondence did not establish correspondence in within-cell response variation.

To calibrate these distances against sampling variation, we compared Qwen–human distances with distances between two matched human samples. Across 18 near/far party-by-tenure cells, mean total-variation distance was 0.561 for Qwen–human and 0.098 for human–human comparisons. The corresponding 95% bootstrap intervals were [0.535, 0.586] and [0.077, 0.119]. Jensen–Shannon divergence was 0.249 versus 0.009, and ordinal Wasserstein distance was 0.721 versus 0.147. For every metric, all 18 Qwen–human cell distances exceeded their matched human–human baselines (Supplementary Fig. S1b–d and Supplementary Table S4).

### 4.2.3 Question-order sensitivity

We then tested whether randomized near/far question order altered the proximity estimates (Fig. 4).

The near-first minus far-first shift in θ was +0.367 for Qwen, −0.100 for Gemma 3, +0.367 for Llama and −0.163 for Mistral. Each remained different from zero after Holm correction within the contrast family. The largest shifts in the overall proximity effect were +0.861 for Llama and +0.897 for Qwen 2.5 7B. The estimated human θ shift was +0.036 (95% CI [−0.169, +0.235]) and did not differ from zero. We did not estimate direct model–human contrasts in order sensitivity, so these within-source tests cannot establish that a model was more order-sensitive than humans. Qwen's primary equivalence result was confined to the pooled θ. Its party-specific contrasts and response distributions remained discrepant, and θ changed with question order.

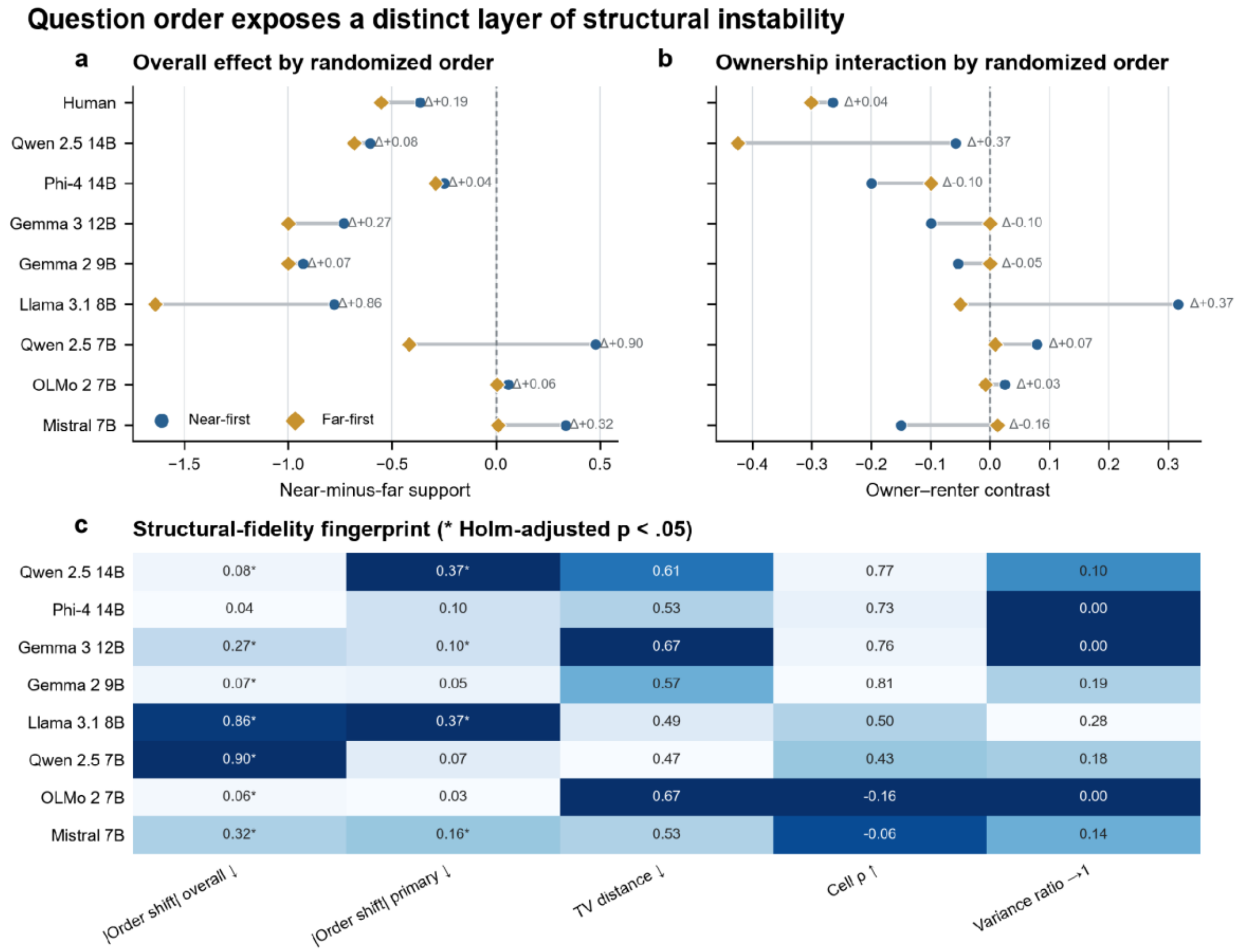

Fig. 4. Randomized question order exposes structural instability not captured by pooled effects. Panels a and b compare near-first and far-first estimates; panel c combines order shifts with distributional and cell-structure diagnostics.

Note. Asterisks mark Holm-adjusted $p < .05$ within the plotted contrast family. Direction and magnitude should be read together; order stability is not interchangeable with distributional fidelity.

### 4.3 Model responses to identity cues and questionnaire context

Qwen completed all 640 cue-source sessions and all 16 required visible profiles. Its profile-equal-weighted support effects were −0.031 in the complete condition, −0.200 in party-only, 0 in ownership-only, and 0 in neither, producing a nonadditive contrast of +0.169 (95% CI [+0.106, +0.236]). Vote and the combined support–vote outcome yielded the same point of contrast, and the perceived-impact composite yielded +0.069. The result supports nonadditive dependence of observable proximity responses on the joint cue configuration. It does not establish a cross-model regularity or an internal fusion mechanism.

Mistral produced strict-complete outputs for 351 of the 640 planned cue-source sessions, including none of the 120 sessions in ownership-only condition. Because this condition was entirely unobserved, the four-condition contrast could not be estimated; its worst-case bounds included negative, zero and positive effects. Restricting the analysis to profiles with valid outputs would have changed the target comparison. Phi and Gemma did not proceed to the full cue-source experiment. In a one-attempt 32-session coverage screen spanning 16 visible profiles and two question orders, Phi and Gemma produced strict-complete outputs in 18/32 and 14/32 sessions, respectively. Each model had at least one profile for which neither question order produced a valid response. We therefore treated these results as incomplete coverage, not as evidence of a zero-cue effect.

Qwen also completed all 270 prompt-framing sessions and all 720 matched-context sessions. Relative to minimal roleplay, the conditional-prediction shift in $\theta$ was −0.036, and the structured-profile shift was −0.185; neither was clear after Holm adjustment. Across the four context conditions, the full-minus-minimal shift in $\theta$ was +0.133, and the nonadditive context interaction in $\theta$ was −0.067, again without a clear adjusted difference. Although none of these $\theta$ shifts was Holm-clear, framing and context shifted the overall near-minus-far support effect: +0.245 under conditional prediction, +0.611 under structured completion, +0.289 for full versus minimal context, and −0.461 for the nonadditive context interaction (all Holm-adjusted $p \leq .006$).

By contrast, Mistral produced incomplete responses that were unevenly distributed across the planned identity groups. It completed 229 of the 270 prompt-framing sessions, but only 59 of the 90 structured-profile sessions met the predefined completion criteria, leaving some identity cells unrepresented. The loss of coverage was greater in the context experiment. Only 171 of the 720 sessions were complete, and just 29 of the 180 matched sets contained valid responses under all four context conditions. Moreover, all 29 complete sets represented Democratic profiles. Consequently, the planned comparison across the three party groups could not be estimated. Because the available responses represented only a restricted subset of the intended population, excluding the incomplete cases would have changed the population addressed by the analysis. We therefore treated nonresponse as evidence about the validity and population coverage of the measurement procedure, rather than as a routine data-cleaning problem (Supplementary Table S5).

## 4.4 Rationale elicitation and representation diagnostics

### 4.4.1 Rationale-first elicitation changed choices

We compared matched direct-choice, choice-first, and rationale-first protocols to determine whether explanation timing altered the measured choice (Table 3 and Fig. 5). Across four focal items, choice-first responses agreed with matched direct choices in 96.4% of Qwen cases, 79.4% of Phi cases, and 98.6% of Gemma cases. When the rationale preceded the choice, the corresponding agreement rates were 79.4%, 70.3% and 64.7%. Rationale-first choices differed from matched direct choices in 20.6% of Qwen cases, 29.7% of Phi cases, and 35.3% of Gemma cases. Mean item scores fell by 0.161 for Qwen and 0.356 for Gemma, whereas the mean Phi shift was −0.011. For Phi, 15.3% of choices became less supportive and 14.4% became more supportive, so the transitions largely cancelled in the mean.

Table 3. Matched reasoning agreement, item-level choice shifts, and the fixed exploratory RSA diagnostic.

| Model | Protocol | Exact agreement | Mean item shift | Fixed RSA [95% CI] | Permutation p |
|---|---|---|---|---|---|
| Qwen 2.5 14B | Choice to rationale | 96.4% | +0.014 | −0.286 [−0.629, 0.000] | .635 |
| Qwen 2.5 14B | Rationale to choice | 79.4% | −0.161 | −0.286 [−0.629, 0.000] | .635 |
| Phi-4 14B | Choice to rationale | 79.4% | −0.044 | 0.000 [−0.114, +0.171] | 1.000 |
| Phi-4 14B | Rationale to choice | 70.3% | −0.011 | 0.000 [−0.114, +0.171] | 1.000 |
| Gemma 3 12B | Choice to rationale | 98.6% | −0.003 | +0.171 [−0.171, +0.400] | .782 |
| Gemma 3 12B | Rationale to choice | 64.7% | −0.356 | +0.171 [−0.171, +0.400] | .782 |

Note. Agreement and mean item shifts compare each protocol with the matched direct choice across four focal items. RSA is repeated across the two protocol rows only for compact presentation; it is one model-level diagnostic.

The dictionary-coded content of the elicited rationales also differed across protocols. When rationales preceded choices, distance references increased by 32 percentage points for Qwen, 34 points for Phi and 14 points for Gemma. Qwen mentioned affordability 15 points less often and neighborhoods 14 points more often, whereas Phi mentioned uncertainty 27 points more often. These frequencies describe elicited text; they do not measure internal reasoning. Because rationale-first elicitation also changed choices, its text and the direct-choice responses came from different measurement protocols. The design did not estimate rationale content as a mediator of direct choices. Further transition and perceived-impact results are reported in Supplementary Section S5, Supplementary Table S6 and Supplementary Figs. S4–S5.

#### 4.4.2 The fixed RSA detected no shared cross-model pattern

We next examined whether hidden-state geometry tracked spatial distance in housing-attitude prompts beyond the structure associated with literal distance tokens. Under the fixed RSA specification, the estimated attitude-specific correlation was negative for Qwen, near zero for Phi and positive for Gemma, but the confidence interval for every model included zero. We therefore found no shared positive pattern across the three models (Supplementary Section S6, Supplementary Table S7 and Supplementary Fig. S6).

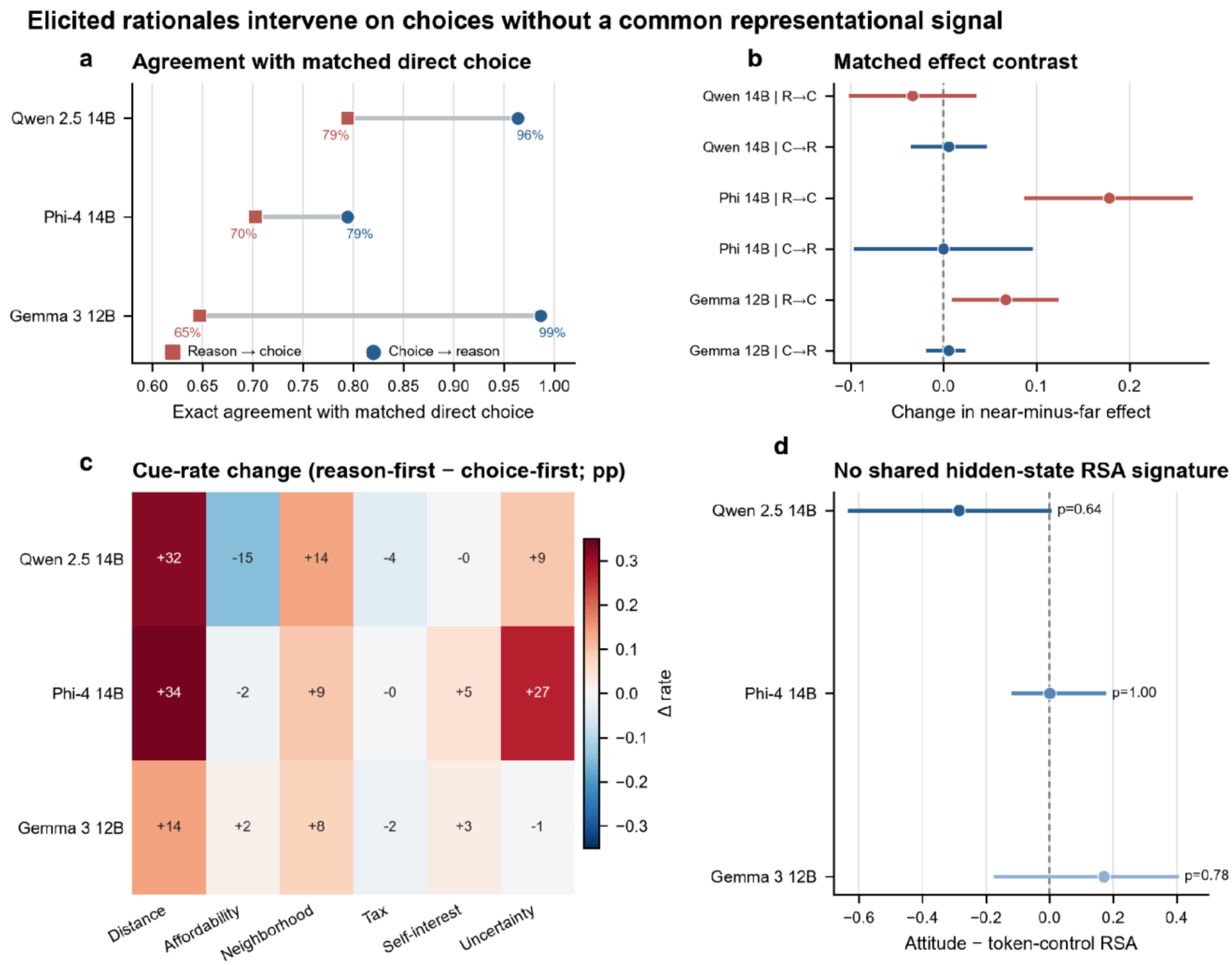

Fig. 5. Rationale elicitation changed choices, while the fixed exploratory RSA yielded no shared positive diagnostic. Panels a–c show matched agreement, effect changes, and cue-rate changes; panel d reports the fixed RSA specification.

Note. R to C denotes rationale first, then choice; C to R denotes choice first, then rationale. RSA intervals use 2,000 bootstrap draws and p-values use 5,000 permutations.

## 5 Discussion

### 5.1 Replicating an effect is not the same as representing a population

Our results reconcile two apparently conflicting findings in recent work. Qwen came closest to the human ownership-conditioned proximity effect and was the only model whose model–human difference met the study's ±0.20 equivalence criterion. Yet it overestimated the overall proximity penalty, compressed within-cell variation and partly cancelled party-specific errors. Agreement on one aggregate estimand can therefore coexist with substantial population mismatch. Phi produced a lower 27-cell RMSE but almost no within-cell response variation. The Gemma models responded strongly to distance but did not reproduce the human ownership gradient. These results show that correspondence on one summary measure is distinct from correspondence with the conditional structure and heterogeneity of a population. This pattern is consistent with prior work showing that LLMs can reproduce some aggregate results of social-science experiments (Ashokkumar et al., 2026; Cui et al., 2025). In some survey settings, model-generated average responses are also close to human baselines (Argyle et al., 2023; Kaiser et al., 2025). However, performance declines once evaluation moves to subgroup-level comparisons. For example, previous work has found substantial misalignment in LLM-generated opinions across 60 US demographic groups, with particularly poor representation of groups such as older and widowed respondents (Santurkar et al., 2023). LLMs also tend to compress response variance and reduce heterogeneity, making it difficult to recover full human response distributions (Bisbee et al., 2024; Wang et al., 2025; Zhang et al., 2025). These findings suggest that finer-grained personas, more careful role construction, and model fine-tuning may improve alignment, but also underscore the need for human validation and rigorous calibration (Lutz et al., 2025; Zhang et al., 2026).

This distinction is important because the human respondents actually rented or owned their homes and reported their own party identities. Tenure may determine residents' perceptions of the project's benefits and risks in the project, while party identity may influence how they weigh the project's local costs against its broader benefits (Hankinson, 2018; Marble and Nall, 2021). Moreover, residents within the same party–tenure group may not hold homogeneous attitudes. Differences in their housing histories, including residential duration, housing stability, and previous exposure to community development, may shape their place attachment and community participation, thereby influencing their level of support for the proposed development (Lewicka, 2011; Manzo and Perkins, 2006). Therefore, tenure and party may organize human responses through several underlying pathways, but the identity labels alone do not establish that the models reproduced those pathways. Because these pathways were not directly measured, they remain plausible explanations rather than established mediators. This distinction clarifies why party-specific errors and compressed within-group variation are substantively important: a model may respond to the supplied tenure and party cues while still misidentifying which groups are most sensitive to project proximity or representing internally diverse groups as unrealistically uniform.

The same distinction clarifies what a behavioral Turing test can and cannot establish. Evaluating whether model responses fall within a human distribution and respond to framing is informative (Mei et al., 2024), but an urban-policy claim also requires the correct conditional structure across groups whose material relations to place differ. Our experiment extends this logic from general behavioral resemblance to a place-based policy response: local validity depends both on where an attitude is elicited and on whether the people exposed to that place are represented with the appropriate heterogeneity (Bollen et al., 2026; Malekzadeh et al., 2025).

### 5.2 The model interface is part of urban-attitude measurement

Randomized question order and input audits show that the interface helps constitute the observed result. A pooled near–far contrast averages over two conversational histories; if the two randomized orders yield different estimates, the pooled number is not a stable property of the model alone. Human survey research offers a useful but limited comparison: earlier questions can change how later questions are understood and which considerations come to mind (Schwarz, 1999). However, as discussed above, human respondents retain their previous experiences and political identities when these are not mentioned, whereas the model sessions only receive these identities through information supplied in the prompt. Context changes and cue removal therefore test the stability of the model-based measurement procedure, not whether the corresponding identities or considerations are absent from human judgment. Prompt framing and preceding context may shift response levels, but they can also determine whether an identity cell produces any valid response. This makes valid completion, profile coverage, and estimability first-class outcomes rather than data-cleaning footnotes. Previous research similarly shows that LLM-generated survey outputs are shaped by prompt design, identity cues and nonresponse mechanisms. This is because models are highly sensitive to prompt structure, labels, example order and tone (Brucks and Toubia, 2025; Tjuatja et al., 2024), and may also treat identity cues as strong conditioning signals while relying on statistical associations learned from training data rather than stable "identity truths". These findings suggest that researchers should aggregate across multiple prompts or full-factorial prompt designs rather than relying on a single prompt and should standardize prompts and maintain fixed version control (Brucks and Toubia, 2025; Tjuatja et al., 2024). Additional safeguards include reducing reliance on a single identity anchor, using interview-style or name-based priming, incorporating information from real interviews and retaining human validation (Lutz et al., 2025; Zhang et al., 2026).

The nonresponse pattern also carries representational consequences. Mistral's successful matched-context pairs came only from Democratic profiles. A complete-case factorial would therefore describe a different target population from the one defined by the design. Likewise, Phi and Gemma cue-coverage criteria failed because profiles and orders were missing, not because the aggregate number of valid strings was merely low. Validity depends on preserving the comparison structure required by the estimand.

### 5.3 Generated justifications do not reveal a common mechanism

The rationale experiment provides a direct warning against treating fluent explanations as passive readouts. Asking for a rationale before the answer changed one-fifth to more than one-third of focal choices, with large model-specific directional effects. It also changed which cues appeared in the text. This is closely related to the distinction between plausible and faithful explanations: a rationale can be coherent and aligned with the final answer while still being produced by an intervention that changed that answer (Turpin et al., 2023). Rationale coding can diagnose sensitivity to response format, but it does not identify the psychological or computational process that produced the choice.

LLM-generated explanations are often post hoc rationalizations rather than transparent records of the underlying reasoning process, and internal representations, probes, or mechanistic analyses alone are insufficient to establish that models share the same mechanism with humans or with one another. This is because chain-of-thought reasoning or self-explanations are often not fully aligned with final answers and may lack independently verifiable validity; in some cases, they are merely texts that appear plausible (Chuang et al., 2026; Randl et al., 2025; Turpin et al., 2023). Explanations should therefore be treated as objects requiring validation rather than as evidence in themselves. Counterfactual, ablation, causal-mediation, and intervention-based tests are needed to assess whether explanations genuinely influence or faithfully reflect model answers (Chuang et al., 2026).

The RSA analysis adds a bounded observation rather than mechanism evidence. Qwen was closest to the human behavioral contrast but had a negative fixed RSA estimate; Gemma had a positive but uncertain estimate despite weak ownership moderation; and Phi was near zero. Because every interval included zero and only three models were evaluated, we do not treat these differences as an established cross-model dissociation. The supported conclusion is narrower: the fixed diagnostic did not recover a shared attitude-specific pattern, so the behavioral result cannot be extended into a common representational-mechanism claim.

### 5.4 Implications for computational urban research and planning

For urban research, LLMs may be useful for pilot testing before human-participant studies, including generating policy scenarios, synthesizing hypotheses, and screening possible aggregate directions of response (Cui et al., 2025; Kaiser et al., 2025). When used to assess public acceptance of planning policies, they may function as scenario simulators, helping researchers rapidly explore policy narratives, generate alternative survey items and test whether a planning proposal might be acceptable at the aggregate level (Kaiser et al., 2025). However, they should not replace resident surveys or community interviews as sources of evidence about subgroup preferences, especially in research involving marginalized groups, spatial inequality, or highly differentiated communities (Bisbee et al., 2024; Wang et al., 2025). In such contexts, LLMs may underestimate the real differences among marginalized populations, sensitive issues, and minority groups; identity-group representation may be flattened, and open-ended responses may become homogenized (Wang et al., 2025; Zhang et al., 2025).

For computational urban research, the validation target should match the intended decision claim. Aggregate calibration may be informative when the task is limited to screening whether a treatment is likely to move an average response, but it is insufficient when outputs are used to characterize residents, neighborhoods, or distributive conflict. The present study did not test policy forecasting on unseen interventions or site-specific decision support, so it does not establish either application. It instead shows why model outputs cannot be treated as substitutes for affected residents solely because one pooled contrast aligns (Bisbee et al., 2024; Bollen et al., 2026; Malekzadeh et al., 2025; Wang et al., 2025).

Together, misaligned subgroup patterns, compressed variation within groups, and selective nonresponse motivate a population-aware approach to reporting urban AI evaluations. Studies should state which population and spatial comparison an output is intended to represent, assess full response distributions rather than means alone, preserve the identity and experimental condition cells required by the estimand, randomize or counterbalance question order, and report condition-specific completion and nonresponse. These practices safeguard the evaluation of urban attitudes generated by models. They do not establish that an untested participation system based on LLMs is ready for planning practice.

### 5.5 Limitations

The study has several principal limitations. First, the human benchmark is one U.S. affordable-housing survey. Although the benchmark respondents had their own observed party and tenure identities, these characteristics may involve multiple experiential and cognitive mediators. Therefore, the design identifies a behavioral contrast conditional on observed identities, not the psychological mechanism producing that contrast. To explore mechanistic differences, further human experiments are needed with matched rationale-order conditions or cognitive measures comparable to the model RSA. In this study, it can be shown only that rationale elicitation changed model choices, but not whether it would similarly affect humans or whether models and humans share an internal mechanism. A human follow-up should test the same direct-choice, choice-first, and rationale-first protocols across party and tenure groups. Moreover, the spatial treatment is experimental distance, not a GIS-based exposure or external validation across real neighborhoods. The composition-weighted sensitivity analysis is conditional on the realized 843-person benchmark and does not establish national representativeness. The Human–Human bootstrap quantifies finite-sample variation conditional on the empirical cell distributions; it does not include recruitment error, survey measurement error, or unobserved population heterogeneity.

Second, model sessions are repeated generations from fixed systems, not sampled residents. Third, the original six-model benchmark used NF4 quantization whereas Phi and Gemma used BF16; we do not interpret cross-system differences as causal effects of family, size, or precision. Fourth, Gemma retained one unresolved main session, and Mistral, Phi, and Gemma exhibited condition-dependent nonresponse in later modules. Fifth, only Qwen and Mistral received the complete three-part input audit, and only Qwen, Phi, and Gemma received matched reasoning and RSA diagnostics. Sixth, the fixed RSA examines hidden-state geometry, not attention weights or causal circuits, and the rationale codebook captures surface content rather than psychological mediation.

Representation analysis remains exploratory. A confirmatory follow-up should prespecify the model checkpoints, target layer, token span, pooling rule, representational metric, controls, and decision threshold; evaluate a held-out stimulus and paraphrase set; and use causal interventions if the goal is mechanism attribution. These requirements describe future validation rather than evidence obtained here. More generally, the eight heterogeneous systems do not support a parameter-count regression, surface rationale codes do not identify psychological mediators, and unplanned model additions or repeated searches across layers would enlarge analytic flexibility without resolving the present research questions.

## 6 Conclusion

Across eight open-weight models we tested, replication of the human proximity response and its moderation by housing tenure varied sharply and did not improve monotonically with model size. Qwen 2.5 14B was the only model to meet the prespecified equivalence criterion for the owner–renter contrast, although it overstated the overall proximity penalty; Phi-4 14B reproduced the direction at a smaller magnitude. The other models either responded strongly to proximity while barely differentiating owners from renters, reversed one of the two effects, or showed little proximity response. Even Qwen's aggregate match did not consistently extend across party cells. Across systems, response variation was compressed, question order altered several estimates, and selective nonresponse prevented some planned comparisons. Requiring a rationale before the choice altered choices in all three models examined, whereas requesting one afterwards left responses nearly unchanged in two of them, and the fixed exploratory RSA yielded no representation pattern shared across those models. Thus, an average effect can look human-like even when subgroup and response patterns do not. For urban-policy research, agreement on an average contrast is therefore not by itself evidence that a model represents resident attitudes. We tested open-weight systems only, and the proprietary models most often proposed for planning use would require the same population-conditional evidence.

**Ethics statement**

The human benchmark was drawn from the control condition of Anderson et al. (2026). Data and replication code for the original study are available through Harvard Dataverse (https://doi.org/10.7910/DVN/DE8RFT). Anderson et al. (2026) report that the original survey was designated exempt by the Dartmouth College Committee for the Protection of Human Subjects. The present study used the archived human data and did not recruit new human participants. To better present our work, the authors used ChatGPT to assist with language polishing and manuscript editing. The authors reviewed, verified, and approved all AI-assisted outputs and remained fully responsible for the content of the manuscript.

## References

Abdin, M., Aneja, J., Behl, H., Bubeck, S., Eldan, R., Gunasekar, S., Harrison, M., Hewett, R.J., Javaheripi, M., Kauffmann, P., Lee, J.R., Lee, Y.T., Li, Y., Liu, W., Mendes, C.C.T., Nguyen, A., Price, E., de Rosa, G., Saarikivi, O., Salim, A., Shah, S., Wang, X., Ward, R., Wu, Y., Yu, D., Zhang, C., Zhang, Y., 2024. Phi-4 Technical Report. https://doi.org/10.48550/ARXIV.2412.08905

Aher, G.V., Arriaga, R.I., Kalai, A.T., 2023. Using Large Language Models to Simulate Multiple Humans and Replicate Human Subject Studies, in: Proceedings of the 40th International Conference on Machine Learning, Proceedings of Machine Learning Research. PMLR, pp. 337–371.

Anderson, C., Briman, E., Ferrin, A., Hampton, C., Malhotra, E., Pandey, S., Robinson, J., Sugerman, L., VanNewkirk, J., Wang, M., Yu, J., Zheng, B., Nyhan, B., 2026. Debunking NIMBY Myths Increases Support for Affordable Housing, Especially Near Respondents' Homes. Journal of Experimental Political Science 13, 178–191. https://doi.org/10.1017/xps.2025.10014

Argyle, L.P., Busby, E.C., Fulda, N., Gubler, J.R., Rytting, C., Wingate, D., 2023. Out of One, Many: Using Language Models to Simulate Human Samples. Political Analysis 31, 337–351. https://doi.org/10.1017/pan.2023.2

Ashokkumar, A., Hewitt, L., Ghezae, I., Willer, R., 2026. Large language models can predict the results of social science experiments. Nature. https://doi.org/10.1038/s41586-026-10742-x

Bisbee, J., Clinton, J.D., Dorff, C., Kenkel, B., Larson, J.M., 2024. Synthetic Replacements for Human Survey Data? The Perils of Large Language Models. Political Analysis 32, 401–416. https://doi.org/10.1017/pan.2024.5

Bollen, P., Higton, J., Sands, M., 2026. Nationally Representative, Locally Misaligned: The Biases of Generative Artificial Intelligence in Neighborhood Perception. Political Analysis 34, 479–487. https://doi.org/10.1017/pan.2025.10022

Brucks, M., Toubia, O., 2025. Prompt architecture induces methodological artifacts in large language models. PLOS One 20, e0319159. https://doi.org/10.1371/journal.pone.0319159

Chuang, Y.-N., Wang, G., Chang, C.-Y., Tang, R., Zhong, S., Yang, F., Wen, A., Du, M., Cai, X., Braverman, V., Hu, X., 2026. FaithLM: Towards Faithful Explanations for Large Language Models. Proceedings of the 19th Conference of the European Chapter of the Association for Computational Linguistics (Volume 1: Long Papers) 3802–3824. https://doi.org/10.18653/v1/2026.eacl-long.177

Cui, Z., Li, N., Zhou, H., 2025. A large-scale replication of scenario-based experiments in psychology and management using large language models. Nat Comput Sci 5, 627–634. https://doi.org/10.1038/s43588-025-00840-7

Dettmers, T., Pagnoni, A., Holtzman, A., Zettlemoyer, L., 2023. QLoRA: Efficient Finetuning of Quantized LLMs, in: Advances in Neural Information Processing Systems 36. Presented at the Advances in Neural Information Processing Systems 36, Neural Information Processing Systems Foundation, Inc. (NeurIPS), New Orleans, Louisiana, USA, pp. 10088–10115. https://doi.org/10.52202/075280-0441

Einstein, K.L., Palmer, M., Glick, D.M., 2019. Who Participates in Local Government? Evidence from Meeting Minutes. Perspect. polit. 17, 28–46. https://doi.org/10.1017/S153759271800213X

Gemma Team, Kamath, A., Ferret, J., Pathak, S., Vieillard, N., Merhej, R., Perrin, S., Matejovicova, T., Ramé, A., Rivière, M., Rouillard, L., Mesnard, T., Cideron, G., Grill, J., Ramos, S., Yvinec, E., Casbon, M., Pot, E., Penchev, I., Liu, G., Visin, F., Kenealy, K., Beyer, L., Zhai, X., Tsitsulin, A., Busa-Fekete, R., Feng, A., Sachdeva, N., Coleman, B., Gao, Y., Mustafa, B., Barr, I., Parisotto, E., Tian, D., Eyal, M., Cherry, C., Peter, J.-T., Sinopalnikov, D., Bhupatiraju, S., Agarwal, R., Kazemi, M., Malkin, D., Kumar, R., Vilar, D., Brusilovsky, I., Luo, J., Steiner, A., Friesen, A., Sharma, Abhanshu, Sharma, Abheesht, Gilady, A.M., Goedeckemeyer, A., Saade, A., Kolesnikov, A., Bendebury, A., Abdagic, A., Vadi, A., György, A., Pinto, A.S., Das, A., Bapna, A., Miech, A., Yang, A., Paterson, A., Shenoy, A., Chakrabarti, A., Piot, B., Wu, B., Shahriari, B., Petrini, B., Chen, C., Lan, C.L., Choquette-Choo, C.A., Carey, C., Brick, C., Deutsch, D., Eisenbud, D., Cattle, D., Cheng, D., Paparas, D., Sreepathihalli, D.S., Reid, D., Tran, D., Zelle, D., Noland, E., Huizenga, E., Kharitonov, E., Liu, F., Amirkhanyan, G., Cameron, G., Hashemi, H., Klimczak-Plucińska, H., Singh, H., Mehta, H., Lehri, H.T., Hazimeh, H., Ballantyne, I., Szpektor, I., Nardini, I., Pouget-Abadie, J., Chan, J., Stanton, J., Wieting, J., Lai, J., Orbay, J., Fernandez, J., Newlan, J., Ji, J., Singh, J., Black, K., Yu, K., Hui, K., Vodrahalli, K., Greff, K., Qiu, L., Valentine, M., Coelho, M., Ritter, M., Hoffman, M., Watson, M., Chaturvedi, M., Moynihan, M., Ma, M., Babar, N., Noy, N., Byrd, N., Roy, N., Momchev, N., Chauhan, N., Bunyan, O., Botarda, P., Caron, P., Rubenstein, P.K., Culliton, P., Schmid, P., Sessa, P.G., Xu, P., Stanczyk, P., Tafti, P., Shivanna, R., Wu, R., Pan, R., Rokni, R., Willoughby, R., Vallu, R., Mullins, R., Jerome, S., Smoot, S., Girgin, S., Iqbal, S., Reddy, S., Sheth, S., Põder, S., Bhatnagar, S., Panyam, S.R., Eiger, S., Zhang, S., Liu, T., Yacovone, T., Liechty, T., Kalra, U., Evci, U., Misra, V., Roseberry, V., Feinberg, V., Kolesnikov, V., Han,

W., Kwon, W., Chen, X., Chow, Y., Zhu, Y., Wei, Z., Egyed, Z., Cotruta, V., Giang, M., Kirk, P., Rao, A., Lo, J., Moreira, E., Martins, L.G., Sanseviero, O., Gonzalez, L., Gleicher, Z., Warkentin, T., Mirrokni, V., Senter, E., Collins, E., Barral, J., Ghahramani, Z., Hadsell, R., Matias, Y., Sculley, D., Petrov, S., Fiedel, N., Shazeer, N., Vinyals, O., Dean, J., Hassabis, D., Kavukcuoglu, K., Farabet, C., Buchatskaya, E., Alayrac, J.-B., Anil, R., Dmitry, Lepikhin, Borgeaud, S., Bachem, O., Joulin, A., Andreev, A., Hardin, C., Dadashi, R., Hussenot, L., 2025. Gemma 3 Technical Report. https://doi.org/10.48550/ARXIV.2503.19786

Gemma Team, Riviere, M., Pathak, S., Sessa, P.G., Hardin, C., Bhupatiraju, S., Hussenot, L., Mesnard, T., Shahriari, B., Ramé, A., Ferret, J., Liu, P., Tafti, P., Friesen, A., Casbon, M., Ramos, S., Kumar, R., Lan, C.L., Jerome, S., Tsitsulin, A., Vieillard, N., Stanczyk, P., Girgin, S., Momchev, N., Hoffman, M., Thakoor, S., Grill, J.-B., Neyshabur, B., Bachem, O., Walton, A., Severyn, A., Parrish, A., Ahmad, A., Hutchison, A., Abdagic, A., Carl, A., Shen, A., Brock, A., Coenen, A., Laforge, A., Paterson, A., Bastian, B., Piot, B., Wu, B., Royal, B., Chen, C., Kumar, C., Perry, C., Welty, C., Choquette-Choo, C.A., Sinopalnikov, D., Weinberger, D., Vijaykumar, D., Rogozińska, D., Herbison, D., Bandy, E., Wang, E., Noland, E., Moreira, E., Senter, E., Eltyshev, E., Visin, F., Rasskin, G., Wei, G., Cameron, G., Martins, G., Hashemi, H., Klimczak-Plucińska, H., Batra, H., Dhand, H., Nardini, I., Mein, J., Zhou, J., Svensson, J., Stanway, J., Chan, J., Zhou, J.P., Carrasqueira, J., Iljazi, J., Becker, J., Fernandez, J., van Amersfoort, J., Gordon, J., Lipschultz, J., Newlan, J., Ji, J., Mohamed, K., Badola, K., Black, K., Millican, K., McDonell, K., Nguyen, K., Sodhia, K., Greene, K., Sjoesund, L.L., Usui, L., Sifre, L., Heuermann, L., Lago, L., McNealus, L., Soares, L.B., Kilpatrick, L., Dixon, L., Martins, L., Reid, M., Singh, M., Iverson, M., Görner, M., Velloso, M., Wirth, M., Davidow, M., Miller, M., Rahtz, M., Watson, M., Risdal, M., Kazemi, M., Moynihan, M., Zhang, M., Kahng, M., Park, M., Rahman, M., Khatwani, M., Dao, N., Bardoliwalla, N., Devanathan, N., Dumai, N., Chauhan, N., Wahltinez, O., Botarda, P., Barnes, P., Barham, P., Michel, P., Jin, P., Georgiev, P., Culliton, P., Kuppala, P., Comanescu, R., Merhej, R., Jana, R., Rokni, R.A., Agarwal, R., Mullins, R., Saadat, S., Carthy, S.M., Cogan, S., Perrin, S., Arnold, S.M.R., Krause, S., Dai, S., Garg, S., Sheth, S., Ronstrom, S., Chan, S., Jordan, T., Yu, T., Eccles, T., Hennigan, T., Kocisky, T., Doshi, T., Jain, V., Yadav, V., Meshram, V., Dharmadhikari, V., Barkley, W., Wei, W., Ye, W., Han, W., Kwon, W., Xu, X., Shen, Z., Gong, Z., Wei, Z., Cotruta, V., Kirk, P., Rao, A., Giang, M., Peran, L., Warkentin, T., Collins, E., Barral, J., Ghahramani, Z., Hadsell, R., Sculley, D., Banks, J., Dragan, A., Petrov, S., Vinyals, O., Dean, J., Hassabis, D., Kavukcuoglu, K., Farabet, C., Buchatskaya, E., Borgeaud, S., Fiedel, N., Joulin, A., Kenealy, K., Dadashi, R., Andreev, A., 2024. Gemma 2: Improving Open Language Models at a Practical Size. https://doi.org/10.48550/ARXIV.2408.00118

Grattafiori, A., Dubey, A., Jauhri, A., Pandey, A., Kadian, A., Al-Dahle, A., Letman, A., Mathur, A., Schelten, A., Vaughan, A., Yang, Amy, Fan, A., Goyal, Anirudh, Hartshorn, A., Yang, Aobo, Mitra, A., Sravankumar, A., Korenev, A., Hinsvark, A., Rao, A., Zhang, A., Rodriguez, A., Gregerson, A., Spataru, A., Roziere, B., Biron, B., Tang, B., Chern, B., Caucheteux, C., Nayak, C., Bi, C., Marra, C., McConnell, C., Keller, C., Touret, C., Wu, C., Wong, C., Ferrer, C.C., Nikolaidis, C., Allonsius, D., Song, D., Pintz, D., Livshits, D., Wyatt, D., Esiobu, D., Choudhary, D., Mahajan, D., Garcia-Olano, D., Perino, D., Hupkes, D., Lakomkin, E., AlBadawy, E., Lobanova, E., Dinan, E., Smith, E.M., Radenovic, F., Guzmán, F., Zhang, F., Synnaeve, G., Lee, G., Anderson, G.L., Thattai, G., Nail, G., Mialon, G., Pang, G., Cucurell, G., Nguyen, H., Korevaar, H., Xu, H., Touvron, H., Zarov, I., Ibarra, I.A., Kloumann, I., Misra, I., Evtimov, I., Zhang, J., Copet, J., Lee, Jaewon, Geffert, J., Vranes, J., Park, Jason, Mahadeokar, J., Shah, J., van der Linde, J., Billock, J., Hong, J., Lee, Jenya, Fu, J., Chi, J., Huang, J., Liu, J., Wang, Jie, Yu, J., Bitton, J., Spisak, J., Park, Jongsoo, Rocca, J., Johnstun, J., Saxe, J., Jia, J., Alwala, K.V., Prasad, K., Upasani, K., Plawiak, K., Li, Ke, Heafield, K., Stone, K., El-Arini, K., Iyer, K., Malik, K., Chiu, K., Bhalla, K., Lakhotia, K., Rantala-Yeary, L., van der Maaten, L., Chen, Lawrence, Tan, L., Jenkins, L., Martin, L., Madaan, L., Malo, L., Blecher, L., Landzaat, L., de Oliveira, L., Muzzi, M., Pasupuleti, M., Singh, M., Paluri, M., Kardas, M., Tsimpoukelli, M., Oldham, M., Rita, M., Pavlova, M., Kambadur, M., Lewis, M., Si, M., Singh, M.K., Hassan, M., Goyal, N., Torabi, N., Bashlykov, N., Bogoychev, N., Chatterji, N., Zhang, N., Duchenne, O., Çelebi, O., Alrassy, P., Zhang, P., Li, P., Vasic, P., Weng, P., Bhargava, P., Dubal, P., Krishnan, P., Koura, P.S., Xu, P., He, Q., Dong, Q., Srinivasan, R., Ganapathy, R., Calderer, R., Cabral, R.S., Stojnic, R., Raileanu, R., Maheswari, R., Girdhar, R., Patel, R., Sauvestre, R., Polidoro, R., Sumbaly, R., Taylor, R., Silva, R., Hou, R., Wang, Rui, Hosseini, S., Chennabasappa, S., Singh, S., Bell, S., Kim, S.S., Edunov, S., Nie, S., Narang, S., Raparthy, S., Shen, S., Wan, S., Bhosale, S., Zhang, Shun, Vandenhende, S., Batra, S., Whitman, S., Sootla, S., Collot, S., Gururangan, S., Borodinsky, S., Herman, T., Fowler, T., Sheasha, T., Georgiou, T., Scialom, T., Speckbacher, T., Mihaylov, T., Xiao, T., Karn, U., Goswami, V., Gupta, V., Ramanathan, V., Kerkez, V., Gonguet, V., Do, V., Vogeti, V., Albiero, V., Petrovic, V., Chu, W., Xiong, W., Fu, W., Meers, W., Martinet, X., Wang, Xiaodong, Wang, Xiaofang, Tan, X.E., Xia, X., Xie, X., Jia, X., Wang, Xuewei, Goldschlag, Y.,

Gaur, Y., Babaei, Y., Wen, Y., Song, Y., Zhang, Yuchen, Li, Yue, Mao, Y., Coudert, Z.D., Yan, Z., Chen, Z., Papakipos, Z., Singh, A., Srivastava, A., Jain, A., Kelsey, A., Shajnfeld, A., Gangidi, A., Victoria, A., Goldstand, A., Menon, A., Sharma, A., Boesenberg, A., Baevski, A., Feinstein, A., Kallet, A., Sangani, A., Teo, A., Yunus, A., Lupu, A., Alvarado, A., Caples, A., Gu, A., Ho, A., Poulton, A., Ryan, A., Ramchandani, A., Dong, A., Franco, A., Goyal, Anuj, Saraf, A., Chowdhury, A., Gabriel, A., Bharambe, A., Eisenman, A., Yazdan, A., James, B., Maurer, B., Leonhardi, B., Huang, B., Loyd, B., De Paola, B., Paranjape, B., Liu, B., Wu, B., Ni, B., Hancock, B., Wasti, B., Spence, B., Stojkovic, B., Gamido, B., Montalvo, B., Parker, C., Burton, C., Mejia, C., Liu, C., Wang, C., Kim, C., Zhou, C., Hu, C., Chu, C.-H., Cai, C., Tindal, C., Feichtenhofer, C., Gao, C., Civin, D., Beaty, D., Kreymer, D., Li, D., Adkins, D., Xu, D., Testuggine, D., David, D., Parikh, D., Liskovich, D., Foss, D., Wang, D., Le, D., Holland, D., Dowling, E., Jamil, E., Montgomery, E., Presani, E., Hahn, E., Wood, E., Le, E.-T., Brinkman, E., Arcaute, E., Dunbar, E., Smothers, E., Sun, F., Kreuk, F., Tian, F., Kokkinos, F., Ozgenel, F., Caggioni, F., Kanayet, F., Seide, F., Florez, G.M., Schwarz, G., Badeer, G., Swee, G., Halpern, G., Herman, G., Sizov, G., Guangyi, Zhang, Lakshminarayanan, G., Inan, H., Shojanazeri, H., Zou, H., Wang, H., Zha, H., Habeeb, H., Rudolph, H., Suk, H., Aspegren, H., Goldman, H., Zhan, H., Damlaj, I., Molybog, I., Tufanov, I., Leontiadis, I., Veliche, I.-E., Gat, I., Weissman, J., Geboski, J., Kohli, J., Lam, J., Asher, J., Gaya, J.-B., Marcus, J., Tang, J., Chan, J., Zhen, J., Reizenstein, J., Teboul, J., Zhong, J., Jin, J., Yang, J., Cummings, J., Carvill, J., Shepard, J., McPhie, J., Torres, J., Ginsburg, J., Wang, Junjie, Wu, K., U, K.H., Saxena, K., Khandelwal, K., Zand, K., Matosich, K., Veeraraghavan, K., Michelena, K., Li, Keqian, Jagadeesh, K., Huang, Kun, Chawla, K., Huang, Kyle, Chen, Lailin, Garg, L., A, L., Silva, L., Bell, L., Zhang, L., Guo, L., Yu, L., Moshkovich, L., Wehrstedt, L., Khabsa, M., Avalani, M., Bhatt, M., Mankus, M., Hasson, M., Lennie, M., Reso, M., Groshev, M., Naumov, M., Lathi, M., Keneally, M., Liu, M., Seltzer, M.L., Valko, M., Restrepo, M., Patel, M., Vyatskov, M., Samvelyan, M., Clark, M., Macey, M., Wang, M., Hermoso, M.J., Metanat, M., Rastegari, M., Bansal, M., Santhanam, N., Parks, N., White, N., Bawa, N., Singhal, N., Egebo, N., Usunier, N., Mehta, N., Laptev, N.P., Dong, N., Cheng, N., Chernoguz, O., Hart, O., Salpekar, O., Kalinli, O., Kent, P., Parekh, P., Saab, P., Balaji, P., Rittner, P., Bontrager, P., Roux, P., Dollar, P., Zvyagina, P., Ratanchandani, P., Yuvraj, P., Liang, Q., Alao, R., Rodriguez, R., Ayub, R., Murthy, R., Nayani, R., Mitra, R., Parthasarathy, R., Li, R., Hogan, R., Battey, R., Wang, Rocky, Howes, R., Rinott, R., Mehta, S., Siby, S., Bondu, S.J., Datta, S., Chugh, S., Hunt, S., Dhillon, S., Sidorov, S., Pan, S., Mahajan, S., Verma, S., Yamamoto, S., Ramaswamy, S., Lindsay, S., Feng, S., Lin, S., Zha, S.C., Patil, S., Shankar, S., Zhang, Shuqiang, Wang, S., Agarwal, S., Sajuyigbe, S., Chintala, S., Max, S., Chen, S., Kehoe, S., Satterfield, S., Govindaprasad, S., Gupta, S., Deng, S., Cho, S., Virk, S., Subramanian, S., Choudhury, S., Goldman, S., Remez, T., Glaser, T., Best, T., Koehler, T., Robinson, T., Li, T., Zhang, T., Matthews, T., Chou, T., Shaked, T., Vontimitta, V., Ajayi, V., Montanez, V., Mohan, V., Kumar, V.S., Mangla, V., Ionescu, V., Poenaru, V., Mihailescu, V.T., Ivanov, V., Li, W., Wang, W., Jiang, W., Bouaziz, W., Constable, W., Tang, X., Wu, Xiaojian, Wang, Xiaolan, Wu, Xilun, Gao, X., Kleinman, Y., Chen, Y., Hu, Y., Jia, Y., Qi, Y., Li, Yenda, Zhang, Yilin, Zhang, Ying, Adi, Y., Nam, Y., Yu, Wang, Zhao, Y., Hao, Y., Qian, Y., Li, Yunlu, He, Y., Rait, Z., DeVito, Z., Rosnbrick, Z., Wen, Z., Yang, Z., Zhao, Z., Ma, Z., 2024. The Llama 3 Herd of Models. https://doi.org/10.48550/ARXIV.2407.21783

Hankinson, M., 2018. When Do Renters Behave Like Homeowners? High Rent, Price Anxiety, and NIMBYism. American Political Science Review 112, 473–493. https://doi.org/10.1017/s0003055418000035

Hu, T., Collier, N., 2024. Quantifying the Persona Effect in LLM Simulations, in: Proceedings of the 62nd Annual Meeting of the Association for Computational Linguistics (Volume 1: Long Papers). Association for Computational Linguistics, pp. 10289–10307. https://doi.org/10.18653/v1/2024.acl-long.554

Innes, J.E., Booher, D.E., 2004. Reframing public participation: strategies for the 21st century. Planning Theory & Practice 5, 419–436. https://doi.org/10.1080/1464935042000293170

Jiang, A.Q., Sablayrolles, A., Mensch, A., Bamford, C., Chaplot, D.S., Casas, D. de las, Bressand, F., Lengyel, G., Lample, G., Saulnier, L., Lavaud, L.R., Lachaux, M.-A., Stock, P., Scao, T.L., Lavril, T., Wang, T., Lacroix, T., Sayed, W.E., 2023. Mistral 7B. https://doi.org/10.48550/ARXIV.2310.06825

Kaiser, C., Kaiser, J., Manewitsch, V., Rau, L., Schallner, R., 2025. Simulating Human Opinions with Large Language Models: Opportunities and Challenges for Personalized Survey Data Modeling, in: Adjunct Proceedings of the 33rd ACM Conference on User Modeling, Adaptation and Personalization. Presented at the UMAP ’25: 33rd ACM Conference on User Modeling, Adaptation and Personalization, ACM, New York City USA, pp. 82–86. https://doi.org/10.1145/3708319.3733685

Kriegeskorte, N., 2008. Representational similarity analysis – connecting the branches of systems neuroscience. Front. Sys. Neurosci. https://doi.org/10.3389/neuro.06.004.2008

Lakens, D., Scheel, A.M., Isager, P.M., 2018. Equivalence Testing for Psychological Research: A Tutorial. Advances in Methods and Practices in Psychological Science 1, 259–269. https://doi.org/10.1177/2515245918770963

Lewicka, M., 2011. Place attachment: How far have we come in the last 40 years? Journal of Environmental Psychology 31, 207–230. https://doi.org/10.1016/j.jenvp.2010.10.001

Lutz, M., Sen, I., Ahnert, G., Rogers, E., Strohmaier, M., 2025. The Prompt Makes the Person(a): A Systematic Evaluation of Sociodemographic Persona Prompting for Large Language Models. Findings of the Association for Computational Linguistics: EMNLP 2025 23212–23237. https://doi.org/10.18653/v1/2025.findings-emnlp.1261

Malekzadeh, M., Willberg, E., Torkko, J., Toivonen, T., 2025. Urban attractiveness according to ChatGPT: Contrasting AI and human insights. Computers, Environment and Urban Systems 117, 102243. https://doi.org/10.1016/j.compenvurbsys.2024.102243

Manzo, L.C., Perkins, D.D., 2006. Finding Common Ground: The Importance of Place Attachment to Community Participation and Planning. Journal of Planning Literature 20, 335–350. https://doi.org/10.1177/0885412205286160

Marble, W., Nall, C., 2021. Where Self-Interest Trumps Ideology: Liberal Homeowners and Local Opposition to Housing Development. The Journal of Politics 83, 1747–1763. https://doi.org/10.1086/711717

Mei, Q., Xie, Y., Yuan, W., Jackson, M.O., 2024. A Turing test of whether AI chatbots are behaviorally similar to humans. Proceedings of the National Academy of Sciences 121. https://doi.org/10.1073/pnas.2313925121

Monkkonen, P., Manville, M., 2019. Opposition to development or opposition to developers? Experimental evidence on attitudes toward new housing. Journal of Urban Affairs 41, 1123–1141. https://doi.org/10.1080/07352166.2019.1623684

OLMo, T., Walsh, P., Soldaini, L., Groeneveld, D., Lo, K., Arora, S., Bhagia, A., Gu, Y., Huang, S., Jordan, M., Lambert, N., Schwenk, D., Tafjord, O., Anderson, T., Atkinson, D., Brahman, F., Clark, C., Dasigi, P., Dziri, N., Ettinger, A., Guerquin, M., Heineman, D., Ivison, H., Koh, P.W., Liu, J., Malik, S., Merrill, W., Miranda, L.J.V., Morrison, J., Murray, T., Nam, C., Poznanski, J., Pyatkin, V., Rangapur, A., Schmitz, M., Skjonsberg, S., Wadden, D., Wilhelm, C., Wilson, M., Zettlemoyer, L., Farhadi, A., Smith, N.A., Hajishirzi, H., 2025. 2 OLMo 2 Furious. https://doi.org/10.48550/ARXIV.2501.00656

Qwen, Yang, A., Yang, B., Zhang, B., Hui, B., Zheng, B., Yu, B., Li, C., Liu, D., Huang, F., Wei, H., Lin, H., Yang, Jian, Tu, J., Zhang, J., Yang, Jianxin, Yang, Jiaxi, Zhou, J., Lin, J., Dang, K., Lu, K., Bao, K., Yang, K., Yu, L., Li, M., Xue, M., Zhang, P., Zhu, Q., Men, R., Lin, R., Li, T., Tang, T., Xia, T., Ren, Xingzhang, Ren, Xuancheng, Fan, Y., Su, Y., Zhang, Y., Wan, Y., Liu, Y., Cui, Z., Zhang, Z., Qiu, Z., 2024. Qwen2.5 Technical Report. https://doi.org/10.48550/ARXIV.2412.15115

Randl, K., Pavlopoulos, J., Henriksson, A., Lindgren, T., 2025. Mind the gap: from plausible to valid self-explanations in large language models. Machine Learning 114. https://doi.org/10.1007/s10994-025-06838-6

Santurkar, S., Durmus, E., Ladhak, F., Lee, C., Liang, P., Hashimoto, T., 2023. Whose Opinions Do Language Models Reflect?, in: Proceedings of the 40th International Conference on Machine Learning, Proceedings of Machine Learning Research. PMLR, pp. 29971–30004.

Schwarz, N., 1999. Self-reports: How the questions shape the answers. American Psychologist 54, 93–105. https://doi.org/10.1037/0003-066X.54.2.93

Tighe, J.R., 2012. How Race and Class Stereotyping Shapes Attitudes Toward Affordable Housing. Housing Studies 27, 962–983. https://doi.org/10.1080/02673037.2012.725831

Tjuatja, L., Chen, V., Wu, T., Talwalkwar, A., Neubig, G., 2024. Do LLMs Exhibit Human-like Response Biases? A Case Study in Survey Design. Transactions of the Association for Computational Linguistics 12, 1011–1026. https://doi.org/10.1162/tacl_a_00685

Turpin, M., Michael, J., Perez, E., Bowman, S., 2023. Language Models Don't Always Say What They Think: Unfaithful Explanations in Chain-of-Thought Prompting, in: Advances in Neural Information Processing Systems. Curran Associates, Inc., pp. 74952–74965. https://doi.org/10.52202/075280-3275

Wang, A., Morgenstern, J., Dickerson, J.P., 2025. Large language models that replace human participants can harmfully misportray and flatten identity groups. Nature Machine Intelligence 7, 400–411. https://doi.org/10.1038/s42256-025-00986-z

Whittemore, A.H., BenDor, T.K., 2019. Reassessing NIMBY: The demographics, politics, and geography of opposition to high-density residential infill. Journal of Urban Affairs 41, 423–442. https://doi.org/10.1080/07352166.2018.1484255

Zhang, J., Liang, X., Deng, A., Bonge, N., Tan, L., Zhang, L., Zarrett, N., 2026. Leveraging interview-informed LLMs to model survey responses: Comparative insights from AI-generated and human data. Journal of Educational Data Mining 18, 1–24. https://doi.org/10.5281/zenodo.18733538

Zhang, S., Xu, J., Alvero, A., 2025. Generative AI Meets Open-Ended Survey Responses: Research Participant Use of AI and Homogenization. Sociological Methods & Research 54, 1197–1242. https://doi.org/10.1177/00491241251327130

## Supplementary Information

## S1. Equivalence-margin sensitivity

### S1.1 Method

The primary analysis classified model–human equivalence for the owner–renter proximity contrast using the frozen 90% interval for the model–human difference and the study-specific ±0.20 support-point margin. As a post hoc transparency analysis, we applied the same frozen intervals to alternative margins of ±0.10, ±0.15 and ±0.25. A model was classified as satisfying a margin only when its complete 90% interval lay within that margin.

### S1.2 Results

The classification depended on the declared tolerance but did not create a general cross-model equivalence result. No model satisfied the ± 0.10 margin. Qwen satisfied ± 0.15, ± 0.20 and ± 0.25. Phi additionally satisfied ± 0.25, whereas the other six models failed all four margins (Supplementary Table S1). The main-text conclusion therefore continues to refer to the prespecified ± 0.20 criterion.

Supplementary Table S1. Equivalence classification under alternative margins.

| Model | Difference | Frozen 90% CI | ±0.10 | ±0.15 | ±0.20 | ±0.25 |
|---|---|---|---|---|---|---|
| Qwen 2.5 14B | 0.044 | [-0.058, +0.139] | No | Yes | Yes | Yes |
| Phi-4 14B | 0.135 | [+0.041, +0.232] | No | No | No | Yes |
| Gemma 3 12B | 0.235 | [+0.144, +0.328] | No | No | No | No |
| Gemma 2 9B | 0.258 | [+0.168, +0.346] | No | No | No | No |
| Llama 3.1 8B | 0.419 | [+0.290, +0.542] | No | No | No | No |
| Qwen 2.5 7B | 0.329 | [+0.206, +0.451] | No | No | No | No |
| OLMo 2 7B | 0.294 | [+0.203, +0.380] | No | No | No | No |
| Mistral 7B | 0.217 | [+0.118, +0.316] | No | No | No | No |

The ±0.20 column denotes the primary study-specific rule; the other margins are post hoc transparency checks. Differences and classifications were calculated from unrounded estimates.

## S2. Human-composition weighting and Human-Human finite-sample baseline

### S2.1 Method

The primary estimand gave equal weight to the three party groups and, within each party, assigned equal weight to mortgage and outright owners when calculating the owner mean. To assess whether this weighting influenced the comparison, we conducted a post hoc sensitivity analysis using the observed party shares in the 843-person Human benchmark and the observed mortgage-versus-outright composition among owners within each party. The resulting composition-weighted contrast therefore reflects the realized composition of this benchmark rather than a nationally representative population, and we applied the same human-derived weights to every model.

Because the statewide item does not form a near–far pair, this finite-sample calibration was restricted to the two local-distance conditions (1/8 mile and 2 miles), yielding 18 party-by-tenure-by-distance cells rather than the 27 cells used for the broader structural comparison. Within each cell, we drew two independent multinomial samples from the empirical Human distribution, using the Human and corresponding model sample sizes. Ten thousand replicates generated matched Human–Human baselines for total variation, Jensen–Shannon divergence and ordinal Wasserstein distance. The excess distance was defined as the Human–model distance minus the matched Human–Human distance. We report descriptive 95% percentile simulation intervals for these quantities; this analysis calibrates finite-sample variation rather than introducing an additional confirmatory test.

**S2.2 Results**

Reweighting the benchmark by its observed composition did not change the qualitative ordering of the models. The Human contrast moved from −0.285 under the registered equal-weight estimand to −0.298, whereas Qwen moved from −0.242 to −0.185. Thus, the absolute Human–Qwen gap increased from 0.044 to 0.114, although Qwen remained the closest model on the primary owner–renter contrast. Because these weights were introduced post hoc, we treat this analysis as a robustness check rather than as a new equivalence test (Supplementary Tables S2 and S3; Supplementary Fig. S1a).

The distributional mismatch remained much larger than the variation expected from matched Human–Human samples. For Qwen, the mean Human–model total variation was 0.561 (95% simulation interval, 0.535–0.586), compared with 0.098 (0.077–0.119) for the Human–Human baseline. The corresponding Jensen–Shannon divergence was 0.249 versus 0.009, and the ordinal Wasserstein distance was 0.721 versus 0.147. Thus, the average Human–model distance exceeded the matched Human–Human baseline for all three metrics (Supplementary Table S4; Supplementary Fig. S1b–d).

Supplementary Table S2. Human party-by-tenure counts and composition weights.

| Party | Tenure | N | Nine-cell weight | Party weight | Owner-type weight |
|---|---|---|---|---|---|
| Democrat | Rent | 112 | 0.133 | 0.332 | |
| Democrat | Mortgage | 102 | 0.121 | 0.332 | 0.607 |
| Democrat | Outright | 66 | 0.078 | 0.332 | 0.393 |
| Republican | Rent | 115 | 0.136 | 0.427 | |
| Republican | Mortgage | 157 | 0.186 | 0.427 | 0.641 |
| Republican | Outright | 88 | 0.104 | 0.427 | 0.359 |
| Independent | Rent | 83 | 0.098 | 0.241 | |
| Independent | Mortgage | 63 | 0.075 | 0.241 | 0.525 |
| Independent | Outright | 57 | 0.068 | 0.241 | 0.475 |

Owner-type weights are defined only for mortgage and outright ownership within party.

Supplementary Table S3. Registered and Human-composition-weighted primary contrasts.

| Source | Registered | Composition-weighted | Weighted minus registered |
|---|---|---|---|
| Human | -0.285 | -0.298 | -0.013 |
| Qwen 2.5 14B | -0.242 | -0.185 | 0.057 |
| Phi-4 14B | -0.150 | -0.138 | 0.012 |
| Gemma 3 12B | -0.050 | -0.040 | 0.010 |
| Gemma 2 9B | -0.027 | -0.027 | 0.001 |
| Mistral 7B | -0.069 | -0.082 | -0.013 |
| OLMo 2 7B | 0.008 | -0.003 | -0.011 |
| Qwen 2.5 7B | 0.044 | 0.007 | -0.037 |
| Llama 3.1 8B | 0.133 | 0.131 | -0.003 |

The weighted analysis is post hoc and does not redefine the primary equivalence rule.

Supplementary Table S4. Human-model distances and matched Human-Human baselines.

| Model | Metric | Human-model [95% CI] | Human-Human [95% CI] | Excess [95% CI] |
|---|---|---|---|---|
| Qwen 2.5 14B | TV | 0.561 [0.535, 0.586] | 0.098 [0.077, 0.119] | 0.463 [0.430, 0.496] |
| Qwen 2.5 14B | Jensen–Shannon (nats) | 0.249 [0.233, 0.265] | 0.009 [0.006, 0.013] | 0.239 [0.223, 0.256] |
| Qwen 2.5 14B | Wasserstein | 0.721 [0.686, 0.758] | 0.147 [0.113, 0.187] | 0.574 [0.522, 0.624] |
| Phi-4 14B | TV | 0.538 [0.513, 0.563] | 0.098 [0.078, 0.119] | 0.441 [0.408, 0.473] |
| Phi-4 14B | Jensen–Shannon (nats) | 0.242 [0.227, 0.258] | 0.010 [0.006, 0.013] | 0.232 [0.217, 0.249] |
| Phi-4 14B | Wasserstein | 0.705 [0.668, 0.742] | 0.148 [0.113, 0.186] | 0.558 [0.504, 0.609] |
| Gemma 3 12B | TV | 0.609 [0.586, 0.631] | 0.098 [0.078, 0.119] | 0.511 [0.480, 0.541] |
| Gemma 3 12B | Jensen–Shannon (nats) | 0.300 [0.284, 0.317] | 0.009 [0.006, 0.013] | 0.291 [0.274, 0.307] |
| Gemma 3 12B | Wasserstein | 0.802 [0.765, 0.841] | 0.147 [0.113, 0.186] | 0.655 [0.601, 0.706] |
| Gemma 2 9B | TV | 0.570 [0.547, 0.594] | 0.097 [0.078, 0.118] | 0.473 [0.441, 0.505] |
| Gemma 2 9B | Jensen–Shannon (nats) | 0.273 [0.257, 0.289] | 0.009 [0.006, 0.013] | 0.263 [0.247, 0.280] |
| Gemma 2 9B | Wasserstein | 0.798 [0.757, 0.840] | 0.147 [0.114, 0.186] | 0.651 [0.595, 0.706] |
| Mistral 7B | TV | 0.531 [0.505, 0.558] | 0.097 [0.078, 0.119] | 0.434 [0.401, 0.467] |
| Mistral 7B | Jensen–Shannon (nats) | 0.237 [0.222, 0.252] | 0.009 [0.006, 0.013] | 0.227 [0.212, 0.243] |
| Mistral 7B | Wasserstein | 0.829 [0.786, 0.871] | 0.147 [0.113, 0.185] | 0.682 [0.625, 0.736] |
| OLMo 2 7B | TV | 0.689 [0.665, 0.712] | 0.098 [0.077, 0.119] | 0.591 [0.559, 0.622] |
| OLMo 2 7B | Jensen–Shannon (nats) | 0.335 [0.318, 0.352] | 0.010 [0.006, 0.013] | 0.325 [0.308, 0.343] |
| OLMo 2 7B | Wasserstein | 0.851 [0.817, 0.884] | 0.148 [0.113, 0.186] | 0.703 [0.652, 0.752] |
| Qwen 2.5 7B | TV | 0.501 [0.472, 0.529] | 0.097 [0.078, 0.119] | 0.403 [0.368, 0.438] |
| Qwen 2.5 7B | Jensen–Shannon (nats) | 0.194 [0.178, 0.210] | 0.009 [0.006, 0.014] | 0.184 [0.168, 0.201] |
| Qwen 2.5 7B | Wasserstein | 0.698 [0.656, 0.741] | 0.147 [0.113, 0.186] | 0.550 [0.493, 0.607] |
| Llama 3.1 8B | TV | 0.522 [0.495, 0.550] | 0.097 [0.078, 0.119] | 0.425 [0.391, 0.458] |
| Llama 3.1 8B | Jensen–Shannon (nats) | 0.208 [0.191, 0.226] | 0.009 [0.006, 0.013] | 0.198 [0.181, 0.216] |
| Llama 3.1 8B | Wasserstein | 0.787 [0.740, 0.835] | 0.147 [0.113, 0.186] | 0.639 [0.579, 0.698] |

Each row summarizes 18 party-by-tenure-by-distance cells and 10,000 draws per cell.

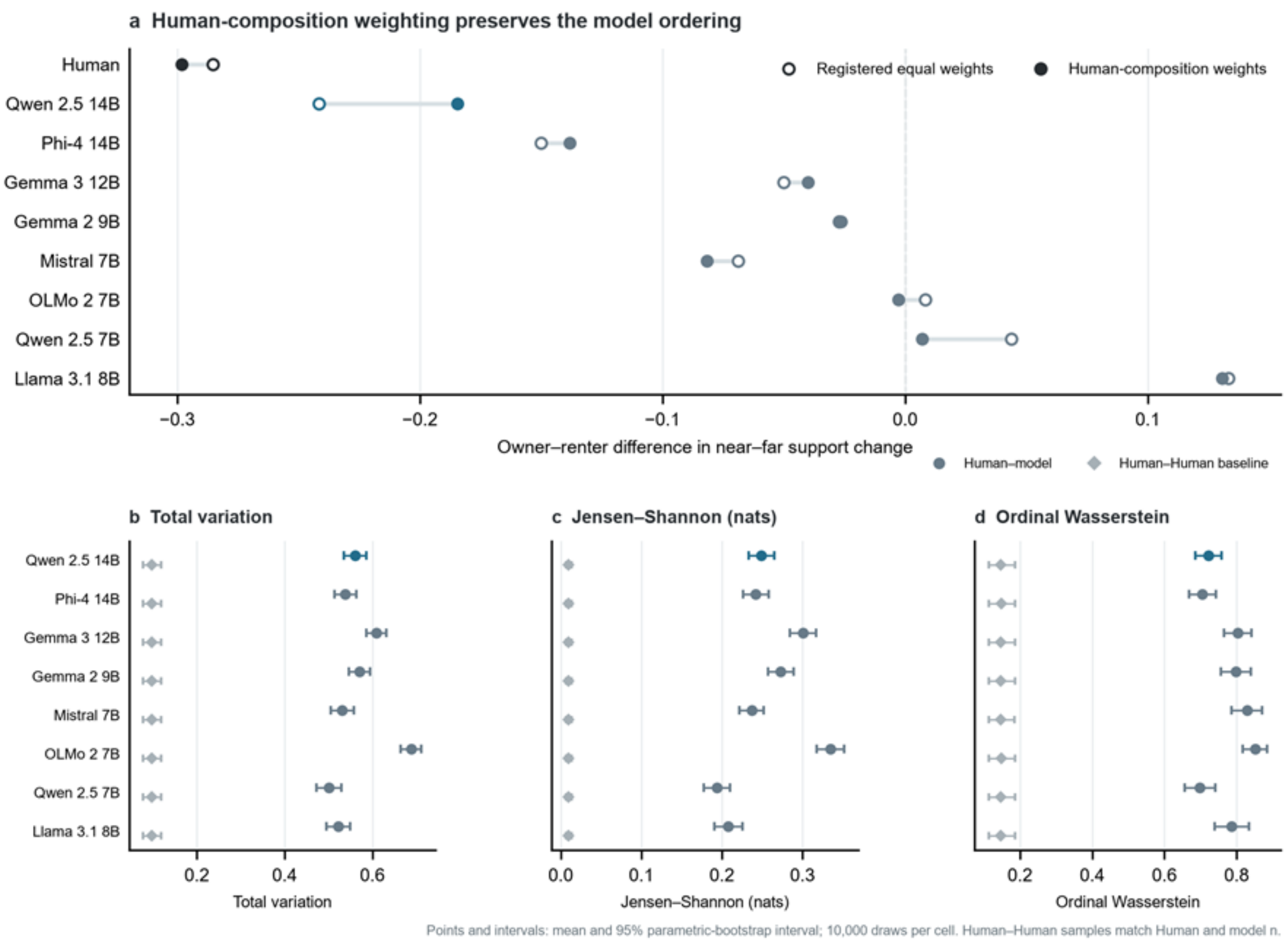

Supplementary Fig. S1. Sensitivity to human population composition and finite-sample distributional variation. a registered equal-weight and human-composition-weighted owner-renter contrast. b-d, mean human-model distances, and matched human-human sampling baselines across 18 party-by-tenure-by-distance distributions. Points and intervals summarize 10,000 multinomial draws per cell.

## S3. Prompt dependence, completion and estimability

### S3.1 Method

For the cue-source ablation, the four-condition contrast required coverage of all 16 visible profiles. For the preceding-context experiment, the registered factorial required 180 matched prompt sets with strict-complete outputs in all four conditions and coverage of all nine party-by-tenure cells. Where missing output prevented point identification, worst-case bounds retained the declared response scale and allowed every unresolved response to take any permitted value.

### S3.2 Results

Qwen completed all cue-source, prompt-framing, and matched-context sessions, so the registered comparisons were estimable. Its cue-source ablation yielded a non-additive support contrast of +0.169, whereas neither the task-framing experiment nor the matched-context experiment produced a Holm-clear shift in the registered proximity contrast.

Mistral completed 351/640 cue-source sessions and 0/120 ownership-only sessions; its four-condition contrast was therefore not estimable. The support bounds (-5.078 to +5.139) included large negative, zero and large positive values. Mistral also completed only 29/180 matched four-condition context sets, all from Democratic profiles. All 32 gate sessions were attempted once for each model, but Phi-4 and Gemma 3 produced strict-complete outputs in only 18 and 14 sessions, respectively; neither proceeded to the full 640-session cue-source design (Supplementary Table S5).

Supplementary Table S5. Completion, estimability, and principal evidence in the input audits.

| System | Module | Strict complete | Estimable? | Principal evidence |
|---|---|---|---|---|
| Qwen 2.5 14B | Prompt framing | 270/270 | Yes | No Holm-clear theta shift |
| Qwen 2.5 14B | Cue-source ablation | 640/640 | Yes | Nonadditive support contrast +0.169 |
| Qwen 2.5 14B | Matched context | 720/720 | Yes | No Holm-clear registered theta shift |
| Mistral 7B | Prompt framing | 229/270 | No | Structured condition missing identity cells |
| Mistral 7B | Cue-source ablation | 351/640 | No | Ownership-only 0/120; bounds cross zero |
| Mistral 7B | Matched context | 171/720 | No | Only 29/180 complete sets; Democratic profiles only |
| Phi-4 14B | Cue-coverage gate | 18/32 | No | Full 640-session design not started |
| Gemma 3 12B | Cue-coverage gate | 14/32 | No | Full 640-session design not started |

Note: The gate comprised 32 sessions spanning 16 profiles and two question orders, with one attempt per session.

A stopped coverage screen or a non-estimable contrast is not a zero-effect result.

## S4. Behavioral failure modes and subgroup errors

### S4.1 Analysis

To clarify how the pooled owner–renter contrast arose, we separated the overall near–far support response from the corresponding responses of owners and renters. We then examined model–Human differences in the near–far support effect within each of the nine party-by-tenure cells. Within each party, mortgage and outright owners contributed equally to the owner response, and ownership moderation was defined as the owner slope minus the renter slope.

### S4.2 Results

Decomposing the responses revealed that several systems reacted strongly to proximity while failing to reproduce the Human tenure contrast. The cell-level error matrix showed why a pooled match can be misleading: subgroup deviations may be attenuated, exaggerated, reversed or cancel one another when averaged across parties and tenure groups (Supplementary Figs. S2 and S3).

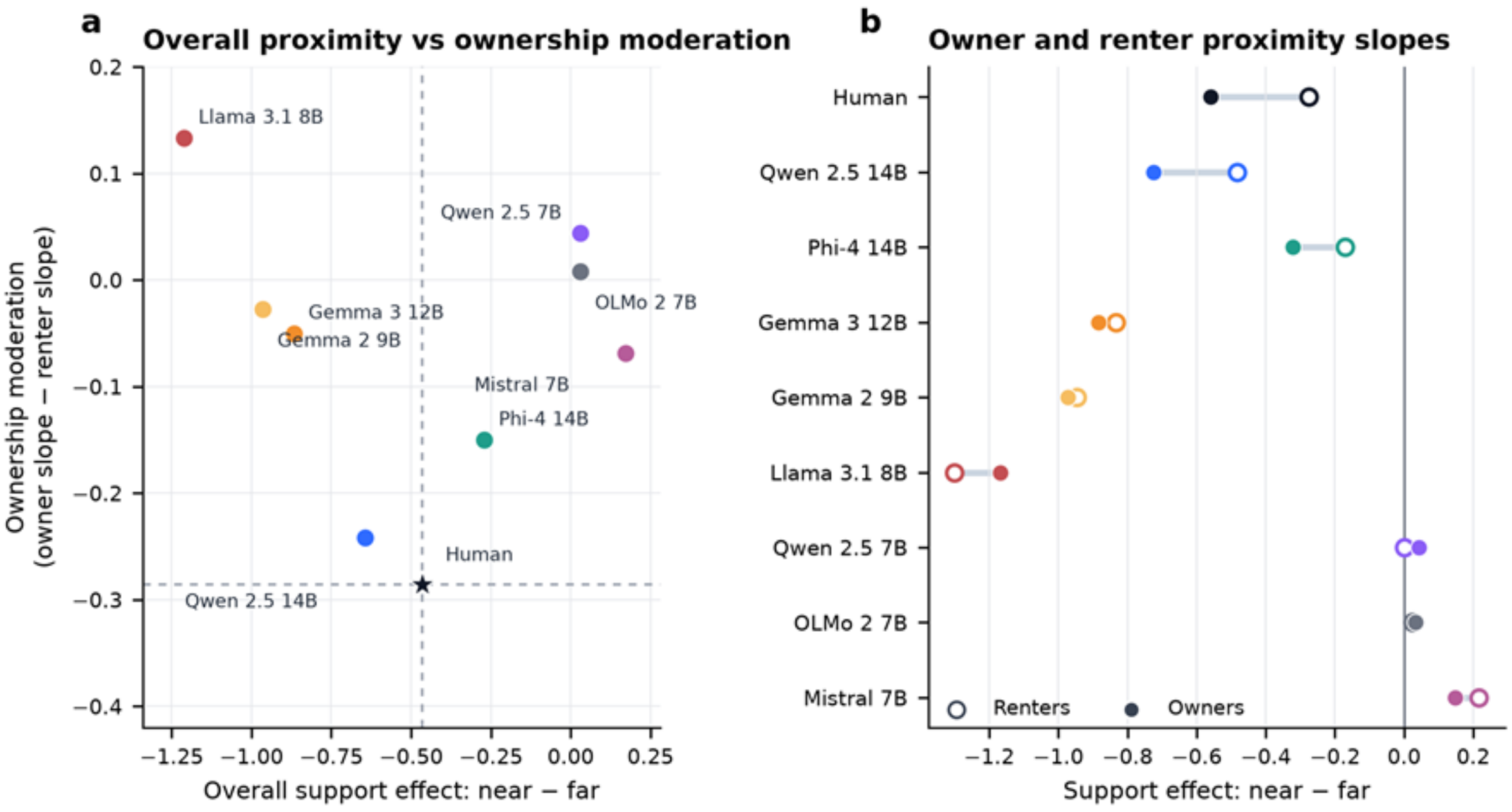

Supplementary Fig. S2. Behavioral failure modes across the Human benchmark and eight models. a, Overall proximity response plotted against ownership moderation; dashed lines mark the Human benchmark. b, Owner and renter near-minus-far support slopes for each model.

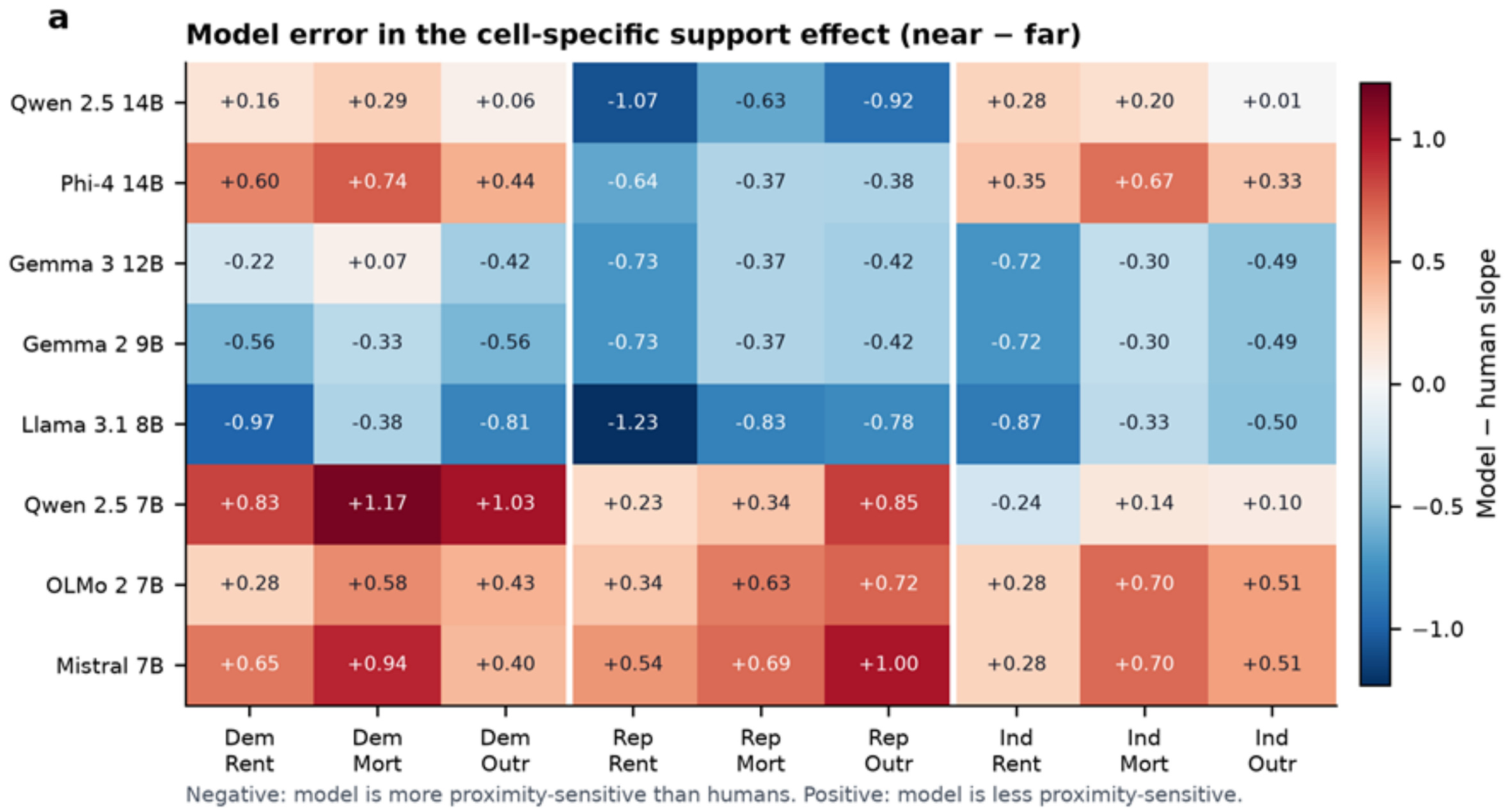

| | Dem Rent | Dem Mort | Dem Outr | Rep Rent | Rep Mort | Rep Outr | Ind Rent | Ind Mort | Ind Outr |
|---|---|---|---|---|---|---|---|---|---|
| Qwen 2.5 14B | +0.16 | +0.29 | +0.06 | -1.07 | -0.63 | -0.92 | +0.28 | +0.20 | +0.01 |
| Phi-4 14B | +0.60 | +0.74 | +0.44 | -0.64 | -0.37 | -0.38 | +0.35 | +0.67 | +0.33 |
| Gemma 3 12B | -0.22 | +0.07 | -0.42 | -0.73 | -0.37 | -0.42 | -0.72 | -0.30 | -0.49 |
| Gemma 2 9B | -0.56 | -0.33 | -0.56 | -0.73 | -0.37 | -0.42 | -0.72 | -0.30 | -0.49 |
| Llama 3.1 8B | -0.97 | -0.38 | -0.81 | -1.23 | -0.83 | -0.78 | -0.87 | -0.33 | -0.50 |
| Qwen 2.5 7B | +0.83 | +1.17 | +1.03 | +0.23 | +0.34 | +0.85 | -0.24 | +0.14 | +0.10 |
| OLMo 2 7B | +0.28 | +0.58 | +0.43 | +0.34 | +0.63 | +0.72 | +0.28 | +0.70 | +0.51 |
| Mistral 7B | +0.65 | +0.94 | +0.40 | +0.54 | +0.69 | +1.00 | +0.28 | +0.70 | +0.51 |

Supplementary Fig. S3. Party-by-tenure cell errors underlying the pooled owner–renter contrast. Cells show model–human differences in near–far support slopes for each party-by-tenure group. Dem, Rep and Ind denote Democrat, Republican, and Independent; Mort and Outr denote mortgage and outright

ownership. Negative values indicate stronger model proximity sensitivity than in the human benchmark, whereas positive values indicate weaker sensitivity.

## S5. Response-format and rationale-content analyses

### S5.1 Method

Because an explanation may influence the choice that follows it, we examined whether response order changed model outputs. We compared Qwen 2.5 14B, Phi-4 14B, and Gemma 3 12B across the same 90 prompt sets. Each set appeared in three forms: a direct choice, a choice followed by a brief explanation, and a brief explanation followed by a choice. We treated direct choices as the reference and compared the two explanation formats across the near and far support and voting items. We also examined six perceived-impact judgements to determine whether protocol sensitivity extended beyond the choices themselves. Agreement, transition direction, and mean response change described format sensitivity.

These comparisons concern observable outputs. They do not identify a hidden reasoning process, and an explanation should not be treated as a transparent report of why the model selected an answer.

### S5.2 Results

Response order mattered, but not equally for all three models. When the choice came first, agreement with matched direct choices was 96.4% for Qwen, 79.4% for Phi and 98.6% for Gemma. Placing the explanation before the choice reduced agreement for every model, to 79.4%, 70.3% and 64.7%, respectively. The especially large decline for Gemma indicates that its final answer was particularly sensitive to the rationale-first protocol, whereas Qwen remained more consistent in that condition. Phi displayed an intermediate pattern. The direction of change also differed across outcomes and perceived-impact domains (Supplementary Table S6; Supplementary Figs. S4 and S5). Together, the findings show that explanations are part of the response protocol, not transparent evidence of a fixed latent rationale.

Supplementary Table S6. Matched response-format transitions relative to direct choice.

| Model | Protocol | Pairs | Less | Same | More | Changed | Mean shift |
|---|---|---|---|---|---|---|---|
| Gemma 3 12B | Choice -> rationale | 360 | 0.8% | 98.6% | 0.6% | 1.4% | -0.003 |
| Phi-4 14B | Choice -> rationale | 360 | 12.5% | 79.4% | 8.1% | 20.6% | -0.044 |
| Qwen 2.5 14B | Choice -> rationale | 360 | 1.1% | 96.4% | 2.5% | 3.6% | 0.014 |
| Gemma 3 12B | Rationale -> choice | 360 | 35.3% | 64.7% | 0.0% | 35.3% | -0.356 |
| Phi-4 14B | Rationale -> choice | 360 | 15.3% | 70.3% | 14.4% | 29.7% | -0.011 |
| Qwen 2.5 14B | Rationale -> choice | 360 | 18.3% | 79.4% | 2.2% | 20.6% | -0.161 |

Rates are calculated across four focal support and vote items within 90 matched prompt sets per model.

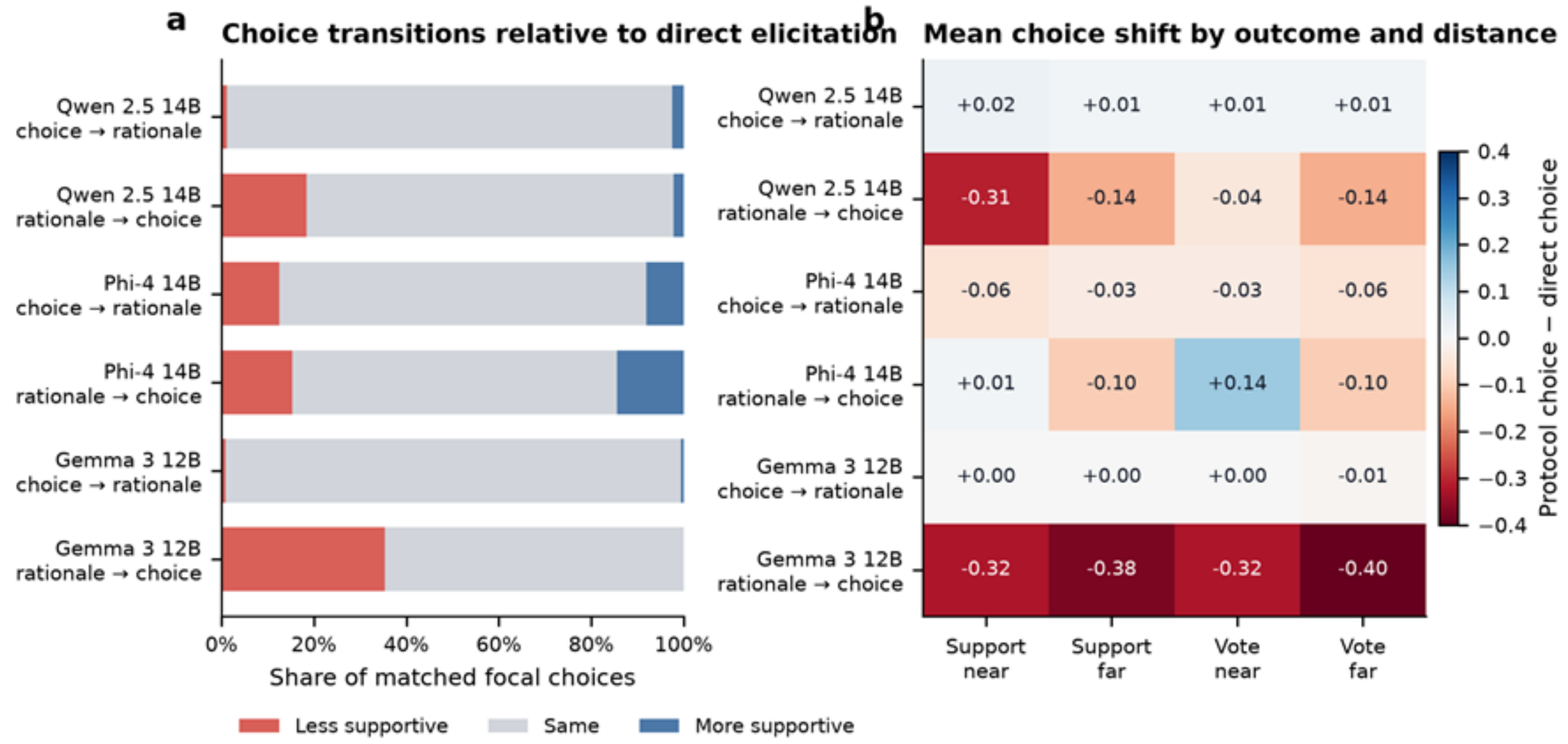

Supplementary Fig. S4. Explanation order altered model choices to different degrees. a, Shares of responses that became less supportive, remained unchanged or became more supportive than the matched

direct choice. b, Mean response shifts by outcome and distance; positive values indicate greater support and negative values indicate lower support.

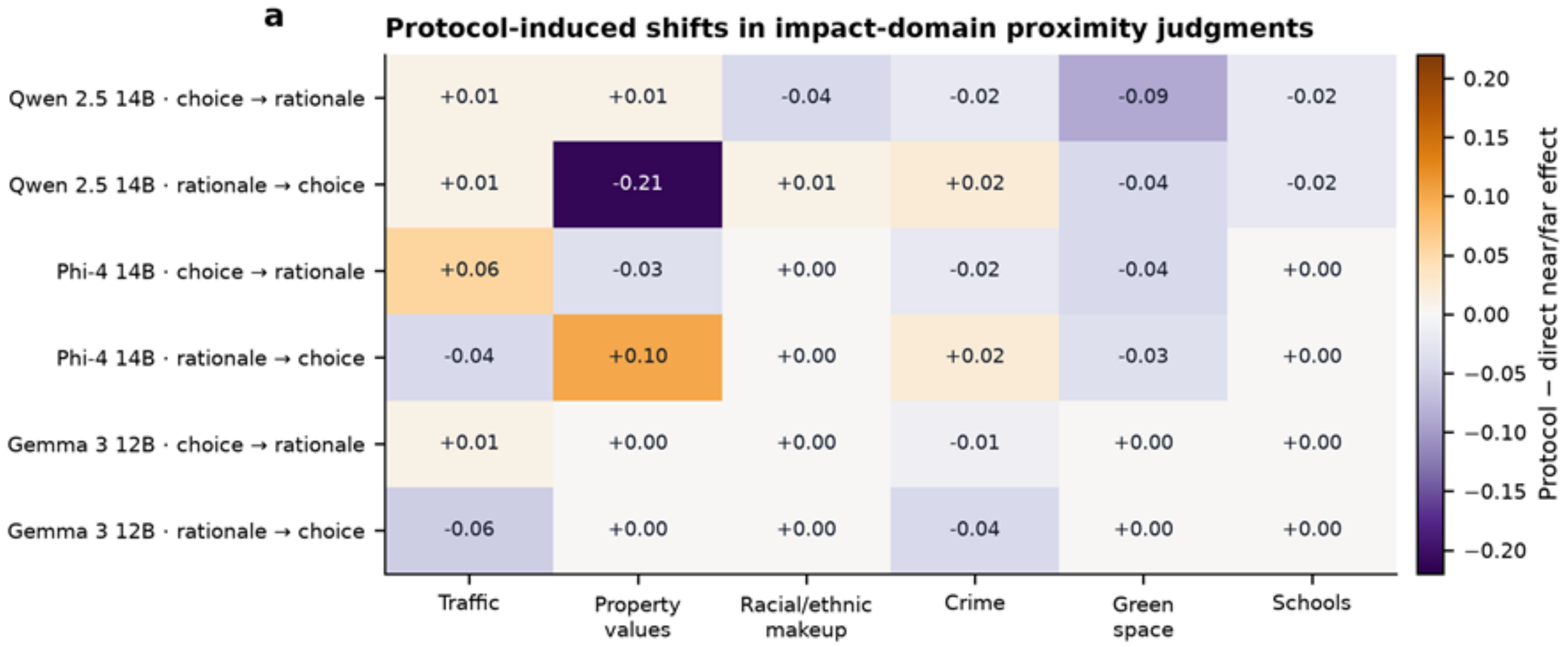

Supplementary Fig. S5. Protocol-induced changes across perceived-impact domains. Cells report protocol-minus-direct differences in near-minus-far impact judgements across perceived-impact domains. Estimates use 90 matched prompt sets per model and response format.

## S6. Layer-wise hidden-state RSA sensitivity

### S6.1 Method

We examined whether the exploratory hidden-state RSA result depended on the selected transformer layer. Using the same 128 prompts and distance-span-mean pooling rule, we compared five positions along each model: 0%, 25%, 50%, 75% and 100% relative depth. This pooling rule averages the hidden states of tokens that express distance. The relative scale allows models with different numbers of layers to be compared, although the same percentage does not denote the same absolute layer number. At each position, uncertainty was summarized with 95% percentile intervals from 2,000 bootstrap resamples. The prespecified 75% layer remained the main reference and was evaluated with 5,000 permutations. Alternative pooling rules served only as robustness checks.

### S6.2 Results

The layer-wise comparison did not reveal a consistent pattern across models. Qwen's largest departure occurred at 75% depth (-0.286), whereas Phi showed only a small departure at 50% (-0.057). Gemma showed positive estimates at both 50% and 75% depth (0.171 at each position), while all remaining point estimates were 0.000. However, every 95% interval included zero, and the location and direction of the deviations differed among models. Together, these results do not support a common depth-stable pattern.

The coarse set of observed coefficients follows from the diagnostic design: four distance centroids yield six unique entries in each representational dissimilarity matrix, and Spearman correlation evaluates the rank ordering of those entries. Taken together, the analysis indicates that any separation between attitude prompts and token controls was model-specific and unstable across depth.

The layer-wise estimates, intervals and permutation values are reported in Supplementary Table S7 and Supplementary Fig. S6.

Supplementary Table S7. Layer-wise attitude-minus-token-control RSA diagnostic.

| **Model** | **Relative depth** | **Estimate** | **[95% CI]** | **Permutation P** | **Primary?** |
|---|---|---|---|---|---|
| Qwen 2.5 14B | 0% | 0.000 | [0.000, 0.000] | 1.000 | No |
| Qwen 2.5 14B | 25% | 0.000 | [-0.229, 0.114] | 1.000 | No |
| Qwen 2.5 14B | 50% | 0.000 | [-0.171, 0.114] | 1.000 | No |
| Qwen 2.5 14B | 75% | -0.286 | [-0.629, 0.000] | 0.635 | Yes |
| Qwen 2.5 14B | 100% | 0.000 | [-0.114, 0.171] | 1.000 | No |
| Phi-4 14B | 0% | 0.000 | [0.000, 0.000] | 1.000 | No |
| Phi-4 14B | 25% | 0.000 | [-0.114, 0.114] | 1.000 | No |
| Phi-4 14B | 50% | -0.057 | [-0.229, 0.171] | 0.919 | No |
| Phi-4 14B | 75% | 0.000 | [-0.114, 0.171] | 1.000 | Yes |
| Phi-4 14B | 100% | 0.000 | [-0.173, 0.400] | 1.000 | No |
| Gemma 3 12B | 0% | 0.000 | [0.000, 0.000] | 1.000 | No |
| Gemma 3 12B | 25% | 0.000 | [-0.114, 0.114] | 1.000 | No |
| Gemma 3 12B | 50% | 0.171 | [-0.499, 0.629] | 0.817 | No |
| Gemma 3 12B | 75% | 0.171 | [-0.171, 0.400] | 0.782 | Yes |
| Gemma 3 12B | 100% | 0.000 | [-0.171, 0.171] | 1.000 | No |

All entries use distance-span-mean pooling. The fixed primary specification is the 75% relative depth.

a Model comparison on a common depth axis

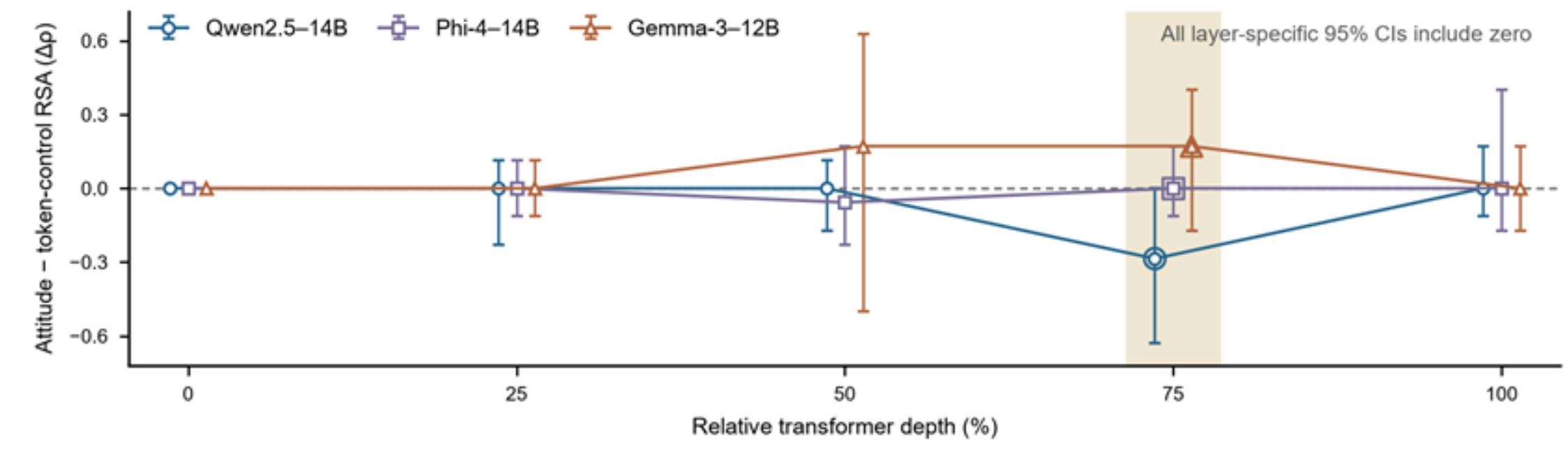

b Fixed 75% layer: decomposition of the model-specific differences

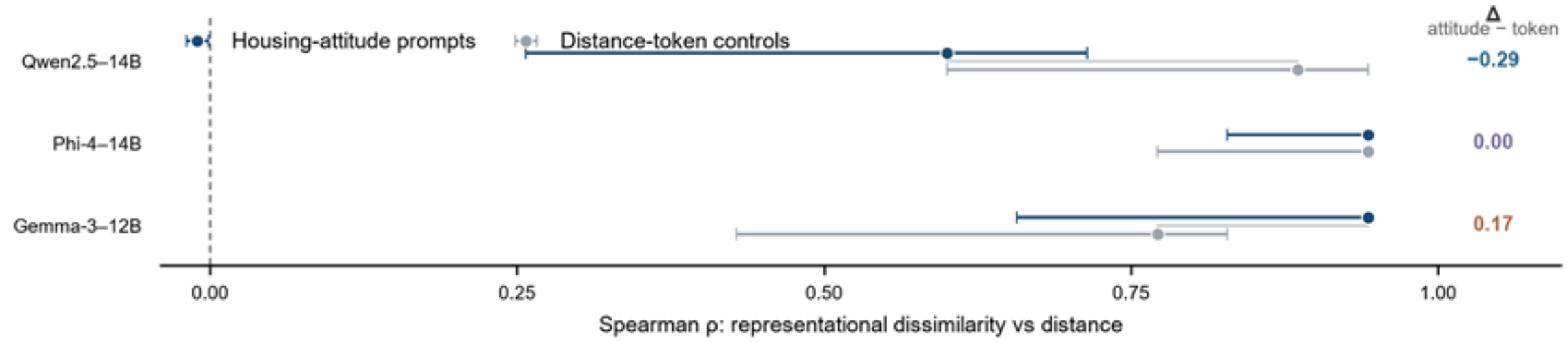

Supplementary Fig. S6. Layer-wise sensitivity of the exploratory hidden-state representational similarity analysis. a, Attitude-minus-token-control RSA estimates across five relative model depths. Points denote estimates, and bars denote 95% bootstrap intervals; shading marks the fixed 75% specification. b, Component RSA correlations at 75% depth.

## S7. Sensitivity to the treatment of initial soft-launch records

The 843-record Human benchmark had been fixed before the provenance check described below. We therefore retained it for the primary analyses and used the 833-record reanalysis to assess sensitivity to the inclusion of the initial soft-launch records.

The source study reports excluding 42 responses from an initial soft launch following a programming error; 10 of these records appeared in our control-condition extraction. We therefore repeated the primary Human analysis after removing these 10 records. The resulting sensitivity sample contained 833 respondents: 108, 101 and 66 Democratic renters, mortgage holders and outright owners; 113, 157 and 88 Republican respondents; and 81, 63 and 56 Independent respondents, respectively.

The Human owner–renter contrast changed from −0.285 (95% CI [−0.385, −0.179]) in the primary benchmark to −0.290 (95% CI [−0.392, −0.188]) in the sensitivity sample. The Qwen 2.5 14B estimate remained −0.242, so the model–Human point-estimate difference changed only from +0.044 to +0.048. Qwen remained the closest model by point estimate.